\documentclass[11pt]{article}

\usepackage[margin=1in]{geometry}
\usepackage[dvips]{graphicx}
\usepackage{amsfonts}
\usepackage{comment}
\usepackage{amsmath}
\usepackage{amssymb}
\usepackage{fancyhdr}
\usepackage{indentfirst}
\usepackage{titlesec}
\usepackage{color}
\usepackage{graphics}
\usepackage{listings}
\usepackage{natbib}
\usepackage{setspace}
\usepackage{bm}
\usepackage{epstopdf}
\usepackage{dsfont}
\usepackage{lscape}
\usepackage{lmodern}
\usepackage{subfig}
\usepackage{float}
\usepackage{fancybox}

\usepackage{endfloat}

\usepackage{hyperref}

\usepackage{setspace}
\newcommand{\defeq}{\mathrel{\mathop:}=}
\numberwithin{equation}{section}

\title{\bf \Large Sparse Vector Autoregressive Modeling}
\date{July 1, 2012}
\author{Richard A. Davis, Pengfei Zang, Tian Zheng \\ Department of statistics, Columbia University}

\begin{document}
\maketitle

\begin{abstract}
The vector autoregressive (VAR) model has been widely used for modeling temporal
dependence in a multivariate time series. For large (and even moderate) dimensions,  the number of
AR coefficients can be prohibitively large, resulting in noisy estimates, unstable predictions and difficult-to-interpret temporal
dependence. To overcome such drawbacks, we propose a 2-stage approach for fitting sparse VAR (sVAR) models in which many of the AR coefficients are zero. The first stage selects non-zero AR coefficients  based on an estimate of the partial spectral coherence (PSC) together with the use of BIC. The PSC is useful for quantifying
the conditional relationship between marginal series in a multivariate process. A refinement second stage is then applied to further reduce the number of parameters. The performance of this 2-stage approach is illustrated with simulation results. The 2-stage approach is also applied to two real data examples: the first is the Google Flu Trends data and the second is a time series of concentration levels of air pollutants.

~\\{\bf Keywords}: vector autoregressive (VAR) model, sparsity, partial spectral coherence (PSC), model selection.
\end{abstract}

\section{Introduction}
The vector autoregressive (VAR) model has been widely used for modeling the temporal dependence structure of a multivariate time series. Unlike univariate time series, the temporal dependence of a multivariate
series consists of not only the serial dependence within each marginal series, but also the interdependence across different marginal series. The VAR model is well suited to describe such temporal dependence structures. However, the conventional VAR model can be saturatedly-parametrized with the number of
AR coefficients prohibitively large for high (and even moderate) dimensional processes. This can result in noisy parameter estimates, unstable predictions and difficult-to-interpret descriptions of the temporal dependence.

To overcome these drawbacks, we propose a 2-stage approach for fitting sparse VAR (sVAR) models in which many of the autoregression (AR) coefficients are zero.
Such sVAR models can enjoy improved efficiency of parameter estimates, better prediction accuracy and more interpretable descriptions of the
temporal dependence structure. In the literature, a class of popular methods for fitting sVAR models is
to re-formulate the VAR model as a penalized regression problem, where the determination of which AR coefficients are zero is equivalent to a variable selection problem in a linear regression setting. One of the most commonly used penalties for the AR coefficients in this context is the Lasso penalty proposed by \citet{Tibshirani1996} and its variants tailored for the VAR modeling purpose, e.g., see \citet{Sosa2005, Hsu2008, Arnold2008, Lozano2009, Haufe2010, Michailidis2010, Song2011}. The Lasso-VAR modeling approach has the advantage of performing model selection and parameter estimation simultaneously. It can also be applied under the ``large-p-small-n'' setting. However, there are also disadvantages in using this approach. First, Lasso has a tendency to over-select the order of the autoregression model and this phenomenon has been reported in various numerical results, e.g., see \citet{Arnold2008, Lozano2009, Michailidis2010}. Second, in applying the Lasso-VAR approach, the VAR model is
re-formulated as a linear regression model, where current values of the time series are treated as the response variable
and lagged values are treated as the explanatory variables. Such a treatment ignores the
temporal dependence in the time series. \citet{Song2011} give a theoretical discussion on the consequences of
applying Lasso directly to the VAR model without taking into account the temporal dependence between
the response and the explanatory variables.

In this paper, we develop a 2-stage approach of fitting sVAR models. The first stage selects non-zero
AR coefficients by screening pairs of distinct marginal series that are conditionally correlated. To compute the
conditional correlation between component series, an estimate of the {\em partial spectral coherence} (PSC)
is used in the first stage. PSC is a tool in frequency-domain time series analysis that can be used to quantify direction-free conditional dependence between component series of a multivariate time series.
An efficient way of computing a non-parametric estimate of PSC is based on results of \citet{Brillinger1981} and \citet{Dahlhaus2000}. In conjunction with the PSC, the {\em Bayesian information criterion} (BIC) is used in the first stage to determine the number of non-zero off-diagonal pairs of AR coefficients. The VAR model fitted in stage 1 may contain spurious non-zero coefficients. To further refine the fitted model, we propose, in stage 2, a screening strategy based on the $t$-ratios of the coefficient estimates as well as BIC.

The remainder of this paper is organized as follows. In Section \ref{section_sVAR}, we review some results on the VAR model for multivariate time series. In Section \ref{section_2stageMethod}, we describe a 2-stage procedure for fitting a sparse VAR model. Connections between our first stage selection procedure with Granger causal models are give in Section \ref{section_2stageMethod_stage1}. In Section \ref{section_simulation}, simulation results are presented to compare the performance of the 2-stage approach against the Lasso-VAR approach. In Section \ref{section_real} the 2-stage approach is applied to fit sVAR models to two real data examples: the first is the Google Flu Trends data (\citet{Ginsberg2009}) and the second is a time series of concentration levels of air pollutants (\citet{Songsiri2009}). Further discussion is contained in Section \ref{section_discussion_conclusion}. Supplementary material is given in the Appendix.

\section{Sparse vector autoregressive models}\label{section_sVAR}
\subsection{Vector autoregressive models (VAR)}\label{section_VARIntro}
Suppose $\{Y_{t}\} = \{(Y_{t,1},Y_{t,2},\ldots,Y_{t,K})^{'}\}$ is a vector autoregressive process of order $p$ (VAR($p$)), which satisfies the recursions,
\begin{equation}
Y_{t} = \mu + \displaystyle\sum_{k=1}^{p}A_{k}Y_{t-k} + Z_{t},~t=0,\pm 1,\ldots, \label{VARequation}
\end{equation}
where $A_{1},\ldots,A_{p}$ are real-valued $K\times K$ matrices of autoregression (AR) coefficients; $\{Z_{t}\}$ are $K$-dimensional iid Gaussian noise with mean $\mathbf{0}$ and non-degenerate covariance matrix $\Sigma_{Z}$. \footnote{In this paper we assume that the VAR($p$) process $\{Y_{t}\}$ is Gaussian. When $\{Y_{t}\}$ is non-Gaussian, the 2-stage model fitting approach can still be applied, where now the Gaussian likelihood is interpreted as a quasi-likelihood.} We further assume that the process $\{Y_{t}\}$ is \textit{causal}, i.e., $\det(I_{K} - \displaystyle\sum_{k=1}^{p}A_{k}z^{k})\ne 0$, for $z \in \mathbb{C}, |z| < 1$, e.g., see \citet{Davis1991} and \citet{Reinsel1997}, which implies that $Z_{t}$ is independent of $Y_{s}$ for $s<t$. Without loss of generality, we also assume that the vector process $\{Y_{t}\}$ has mean $\mathbf{0}$, i.e., $\mu=\mathbf{0}$ in \eqref{VARequation}.

\subsection{Sparse vector autoregressive models (sVAR)}\label{section_sVARIntro}
The temporal dependence structure of the VAR model \eqref{VARequation} is characterized by the AR coefficient matrices $A_{1},\ldots, A_{p}$. Based on $T$ observations $Y_{1},\ldots,Y_{T}$ from the VAR model, we want to estimate these AR matrices. However, a VAR($p$) model, when fully-parametrized, has $K^2p$ AR parameters that need to be estimated. For large (and even moderate) dimension $K$, the number of parameters can be prohibitively large, resulting in noisy estimates, unstable predictions and difficult-to-interpret descriptions of the temporal dependence. It is also generally believed that, for most applications, the true model of the series is sparse, i.e., the number of non-zero coefficients is small. Therefore it is preferable to fit a {\em sparse} VAR (sVAR) model in which many of its AR parameters are zero. In this paper we develop a 2-stage approach of fitting sVAR models. The first stage selects non-zero AR coefficients by screening pairs of distinct marginal series that are conditionally correlated. To compute direction-free
conditional correlation between components in the time series, we use tools from the frequency-domain,
specifically the {\em partial spectral coherence} (PSC). Below we introduce the basic properties related to PSC.

Let $\{Y_{t,i}\}$ and $\{Y_{t,j}\}$ ($i\ne j$) denote two distinct marginal series of the process $\{Y_{t}\}$, and $\{Y_{t,-ij}\}$ denote the remaining $(K-2)$-dimensional process. To compute the conditional correlation between two time series $\{Y_{t,i}\}$ and $\{Y_{t,j}\}$, we need to adjust for the linear effect from the remaining marginal series $\{Y_{t,-ij}\}$. The removal of the linear effect of $\{Y_{t,-ij}\}$ from each of $\{Y_{t,i}\}$ and $\{Y_{t,j}\}$ can be achieved by using results of linear filters, e.g., see \citet{Brillinger1981} and \citet{Dahlhaus2000}. Specifically, the optimal linear filter for removing the linear effect of $\{Y_{t,-ij}\}$ from $\{Y_{t,i}\}$ is given by the set of $(K-2)$-dimensional constant vectors that minimizes the expected squared error of filtering,
\begin{equation}
\{D_{k,i}^{opt} \in \mathbb{R}^{K-2}, k\in \mathbb{Z}\} =
\underset{\{D_{k,i}, k\in \mathbb{Z}\}}{\operatorname{argmin}} \mathbf{E}(Y_{t,i} - \displaystyle\sum_{k=-\infty}^{\infty}D_{k,i}Y_{t-k,-ij})^2. \label{optFilter}
\end{equation}
The {\em residual series} from the optimal linear filter is defined as,
\begin{equation*}
\varepsilon_{t,i} \defeq Y_{t,i} - \displaystyle\sum_{k=-\infty}^{\infty}D_{k,i}^{opt}Y_{t-k,-ij}.
\end{equation*}
Similarly, we use $\{D_{k,j}^{opt} \in \mathbb{R}^{K-2}, k\in \mathbb{Z}\}$ and $\{\varepsilon_{t,j}\}$ to denote the optimal linear filter and the corresponding residual series for another marginal series $\{Y_{t,j}\}$. Then the conditional correlation between $\{Y_{t,i}\}$ and $\{Y_{t,j}\}$ is characterized by the correlation between the two residual series $\{\varepsilon_{t,i}\}$ and $\{\varepsilon_{t,j}\}$. In particular, two distinct marginal series $\{Y_{t,i}\}$ and $\{Y_{t,j}\}$  are {\em conditionally uncorrelated} after removing the linear effect of $\{Y_{t,-ij}\}$ if and only if their residual series $\{\varepsilon_{t,i}\}$ and $\{\varepsilon_{t,j}\}$ are uncorrelated at all lags, i.e., $\mathrm{cor}(\varepsilon_{t+k,i},\varepsilon_{t,j})=0$, for $k \in \mathbb{Z}$. In the frequency domain, $\{\varepsilon_{t,i}\}$ and $\{\varepsilon_{t,j}\}$ are uncorrelated at all lags is equivalent to the cross-spectral density of the two residual series, denoted by $f^{\varepsilon}_{ij}(\omega)$, is zero at all frequencies $\omega$. Here the residual cross-spectral density is defined by,
\begin{equation}
f^{\varepsilon}_{ij}(\omega) \defeq \frac{1}{2\pi}\displaystyle\sum_{k=-\infty}^{\infty}\gamma^{\varepsilon}_{ij}(k)\mathrm{e}^{-\mathrm{i} k\omega},~ \omega \in (-\pi, \pi], \label{resid_crossSpec}
\end{equation}
where $\gamma^{\varepsilon}_{ij}(k) \defeq \mathrm{cov}(\varepsilon_{t+k,i}, \varepsilon_{t,j})$. The cross-spectral density $f^{\varepsilon}_{ij}(\omega)$ reflects the conditional (or partial) correlation between the two corresponding marginal series $\{Y_{t,i}\}$ and $\{Y_{t,j}\}$, given $\{Y_{t,-ij}\}$. This observation leads to the definition of {\em partial spectral coherence} (PSC), e.g., see \citet{Brillinger1981, Davis1991}, between two distinct marginal series $\{Y_{t,i}\}$ and $\{Y_{t,j}\}$, which is defined as the scaled cross-spectral density between the two residual series $\{\varepsilon_{t,i}\}$ and $\{\varepsilon_{t,j}\}$, i.e.,
\begin{equation}
\mathrm{PSC}_{ij}(\omega) \defeq \frac{f^{\varepsilon}_{ij}(\omega)}{\sqrt{f^{\varepsilon}_{ii}(\omega) f^{\varepsilon}_{jj}(\omega)}}, ~ \omega \in (-\pi,\pi]. \label{PSCdefi}
\end{equation}
\citet{Brillinger1981} showed that the cross-spectral density  $f^{\varepsilon}_{ij}(\omega)$ can be computed from the spectral density $f^{Y}(\omega)$ of the process $\{Y_{t}\}$ via,
\begin{equation}
f^{\varepsilon}_{ij}(\omega) =  f^{Y}_{ii}(\omega) - f^{Y}_{i,-ij}(\omega)f^{Y}_{-ij,-ij}
(\omega)^{-1}f^{Y}_{-ij,j}(\omega), \label{residCrossSpec_david}
\end{equation}
which involves inverting a $(K-2)\times (K-2)$ dimensional matrix, i.e., $f^{Y}_{-ij,-ij}
(\omega)^{-1}$. Using \eqref{residCrossSpec_david} to compute the PSCs for all pairs of distinct marginal series of $\{Y_{t}\}$ requires $\binom{K}{2}$ such matrix inversions, which can be computationally challenging for a large dimension $K$. \citet{Dahlhaus2000} proposed a more efficient method to simultaneously compute the PSCs for all $\binom{K}{2}$ pairs through the inverse of the spectral density matrix, which is defined as $g^{Y}(\omega) \defeq f^{Y}(\omega)^{-1}$: Let $g^{Y}_{ii}(\omega)$, $g^{Y}_{jj}(\omega)$ and $g^{Y}_{ij}(\omega)$ denote the $i$th diagonal, the $j$th diagonal
and the $(i,j)$th entry of $g^{Y}(\omega)$, respectively; Then the partial spectral coherence between $\{Y_{t,i}\}$ and $\{Y_{t,j}\}$  can be computed as follows,
\begin{equation}
\mathrm{PSC}_{ij}(\omega) = -\frac{g^{Y}_{ij}(\omega)}{\sqrt{g^{Y}_{ii}(\omega)g^{Y}_{jj}(\omega)}}, ~\omega \in (-\pi,\pi]. \label{inverseSpecDen}
\end{equation}
The computation of all $\binom{K}{2}$ PSCs using \eqref{inverseSpecDen} requires only one matrix inversion of the $K\times K$ dimensional matrix $f^{Y}(\omega)$. It then follows that,
\begin{eqnarray}\label{condUncorToZeroPSC}
&& \{Y_{t,i}\} \mbox{ and } \{Y_{t,j}\}~(i \ne j)\mbox{ are conditionally uncorrelated } \\
&&\mbox{iff}~g^{Y}_{ij}(\omega) = 0, \mbox{ for all } \omega \in (-\pi,\pi]. \nonumber
\end{eqnarray}
In other words, the inverse spectral density matrix $g^{Y}(\omega)$
encodes the pairwise conditional correlation between the component series of $\{Y_{t}\}$. This generalizes the problem of {\em covariance selection} in which independent samples are available, e.g., see \citet{Dempster1972, Friedman2008}. Covariance selection is concerned about the conditional relationship between
dimensions of a multivariate Gaussian distribution by locating zero entries in the inverse covariance matrix. For example, suppose $X=(X_{1},\ldots,X_{K})^{'}$ follows a $K$-dimensional Gaussian $N(0, \Sigma_{X})$. It is known that two distinct dimensions, say $X_{i}$ and $X_{j}$ ($i\ne j$), are conditionally independent given the other $(K-2)$ dimensions $X_{-ij}$, if and only if the $(i,j)$th entry in the inverse covariance matrix $\Sigma_{X}^{-1}$ is zero, i.e.,
\begin{equation}
X_{i} \mbox{ and } X_{j}~(i \ne j)\mbox{ are conditionally independent}~\mbox{iff}~\Sigma_{X}^{-1}(i,j)=0. \label{condIndeGau}
\end{equation}
If the process $\{Y_{t}\}$ were independent replications of a Gaussian distribution N(0,~$\Sigma_{Y}$),
then its spectral density matrix $f^{Y}(\omega)=\Sigma_{Y}$ remains constant over $\omega \in (-\pi,\pi]$ and \eqref{condUncorToZeroPSC} becomes,
\begin{equation}
\small
\{Y_{t,i}\} \mbox{ and } \{Y_{t,j}\}~(i \ne j)\mbox{ are conditionally uncorrelated}~\mbox{iff}~\Sigma_{Y}^{-1}(i,j) = 0,
\normalsize
\end{equation}
which coincides with \eqref{condIndeGau}. Therefore selection of conditionally uncorrelated series using the inverse of spectral density contains the covariance selection problem as a special case.

\section{A 2-stage approach of fitting sVAR models} \label{section_2stageMethod}
In this section, we develop a 2-stage approach of fitting sVAR models. The first stage of the approach takes advantage of \eqref{condUncorToZeroPSC} and screens out the pairs of marginal series that are conditionally uncorrelated. For such pairs we set the corresponding AR coefficients to zero for each lag. However, the model fitted in stage 1 may still contain spurious non-zero AR coefficient estimates. To address this possibility, a second stage is used to refine the model further.

\subsection{Stage 1: selection}\label{section_2stageMethod_stage1}
As we have shown in Section \ref{section_sVARIntro}, a zero PSC indicates that the two corresponding marginal series are conditionally uncorrelated. In the first stage of our approach, we use the information of pairwise conditional uncorrelation to reduce the complexity of the VAR model. In particular, we propose to set the AR coefficients between two conditionally uncorrelated marginal series to zero, i.e.,
\begin{eqnarray}\label{groupARzero}
&& A_{k}(i,j)=A_{k}(j,i)=0~(i\ne j, k=1,\ldots,p) \\
&&\mbox{if}~\{Y_{t,i}\} \mbox{ and } \{Y_{t,j}\} \mbox{ are conditionally uncorrelated}, \nonumber
\end{eqnarray}
where the latter is equivalent to $\mathrm{PSC}_{ij}(\omega)=0$ for $\omega \in (-\pi, \pi]$. From \eqref{groupARzero} we can see that the modeling interest of the first stage is whether or not the AR coefficients belonging to a pair of marginal series at all lags are selected, rather than the selection of an individual AR coefficient. We point out that our proposed connection from zero PSCs to zero AR coefficients, as described by \eqref{groupARzero}, may not be exact for some examples. However, numerical results suggest that our 2-stage approach is still able to achieve well-fitted sVAR models for such examples. We will return to this point in Section \ref{section_discussion_conclusion}.

In order to set a group of AR coefficients to zero as in \eqref{groupARzero}, we need to find the pairs of marginal series for which the PSC is identically zero. Due to sampling variability, however, the estimated PSC, denoted by $\hat{\mathrm{PSC}}_{ij}(\omega)$ for series $\{Y_{t,i}\}$ and $\{Y_{t,j}\}$, will not be exactly zero even when the two corresponding marginal series are conditionally uncorrelated.  In other words, we need to rank the estimated PSC based on their evidence to be non-zero and decide a cutoff point that separates non-zero PSC from zero PSC. Since the estimate $\hat{\mathrm{PSC}}_{ij}(\omega)$ depends on the frequency $\omega$, we need a quantity to summarize its departure from zero over different frequencies.  As in \citet{Dahlhaus2000, Dahlhaus1997}, we use the supremum of the squared modulus of the estimated PSC, i.e.,
\begin{align}\label{supPSCstat}
\hat{S}_{ij} \defeq \underset{\omega}{\operatorname{sup}} |\hat{\mathrm{PSC}}_{ij}(\omega)|^2,
\end{align}
as the summary statistic, where the supremum is taken over the Fourier frequencies $\{2\pi k/T: k=1,\ldots,T\}$. A large value of $\hat{S}_{ij}$ indicates that the two marginal series are likely to be conditionally correlated. Therefore we can create a sequence $\mathbf{Q}_{1}$ of the $\binom{K}{2}$ pairs of distinct marginal series by ranking each pair's summary statistic \eqref{supPSCstat} from highest to lowest. This sequence $\mathbf{Q}_{1}$ prioritizes the way in which non-zero coefficients are added into the VAR model. Based on the sequence $\mathbf{Q}_{1}$, we need two parameters to fully specify the VAR model: the order of autoregression $p$ and the number of {\em top} pairs in $\mathbf{Q}_{1}$, denoted by $M$, that are selected into the VAR model. For the $\frac{(K-1)K}{2}-M$ pairs not selected, their corresponding groups of AR coefficients are set to zero. The two parameters $(p, M)$ control the complexity of the VAR model as the number of non-zero AR coefficients is $(K+2M)p$. We use the BIC, see \citet{Schwarz1978}, to simultaneously choose the values of these two parameters. The BIC is computed as,
\begin{equation}\label{BICsVAR}
\mathrm{BIC}(p,M) = -2\log L(\hat{A}_1,\ldots,\hat{A}_p) + \log{T}\cdot(K+2M)p,
\end{equation}
where $L(\hat{A}_1,\ldots,\hat{A}_p)$ is the maximized likelihood of the VAR model.
To compute the maximized likelihood $L(\hat{A}_1,\ldots,\hat{A}_p)$, we use results on the constrained maximum likelihood estimation of VAR models as given in \citet{Helmut1991}. Details of this estimation procedure can be found Appendix \ref{sVARest}.

Restricting the two parameters $p$ and $M$ to take values in pre-specified ranges $\mathbb{P}$ and $\mathbb{M}$, respectively, the steps of {\bf stage 1} can be summarized as follows.
\begin{center}
\Ovalbox{\setlength{\itemsep}{0pt}
\begin{minipage}{0.97\textwidth}
\begin{center}
{\bf Stage 1}
\end{center}
\begin{itemize}\label{sVAR1stage}
\item[1.] Estimate the $\mathrm{PSC}$ for all $K(K-1)/2$ pairs of distinct marginal series by inverting a non-parametric estimate of the spectral density matrix \footnotemark\ and applying equation \eqref{inverseSpecDen}.
\item[2.] Construct a sequence $\mathbf{Q}_{1}$ of the $K(K-1)/2$ pairs of distinct marginal series by ranking each pair's summary statistic $\hat{S}_{ij}$ \eqref{supPSCstat} from highest to lowest.
\item[3.] For each $(p,M) \in \mathbb{P}\times \mathbb{M}$, set the order of autoregression to $p$ and select the top $M$ pairs in the sequence $\mathbf{Q}_{1}$ into the VAR model, which specifies the parameter constraint on the AR coefficients. Conduct parameter estimation under this constraint using the results in Appendix \ref{sVARest} and compute the corresponding $\mathrm{BIC}(p,M)$ according to equation \eqref{BICsVAR}.
\item[4.] Choose $(\tilde{p}, \tilde{M})$ that gives the minimum BIC value over $\mathbb{P}\times \mathbb{M}$.
\end{itemize}
\end{minipage}}
\end{center}
\footnotetext{In this paper we use the periodogram smoothed by a modified Daniell kernel, e.g., see \citet{Davis1991}, as the non-parametric estimate of the spectral density. Alternative spectral density estimates, such as the shrinkage estimate proposed by \citet{Bohm2009}, can also be adopted.}

The model obtained in the first stage contains $(K+2\tilde{M})\tilde{p}$ non-zero AR coefficients. If only a small proportion of the pairs of marginal series are selected, i.e., $\tilde{M} << K(K-1)/2$, $(K+2\tilde{M})\tilde{p}$ can be much smaller than $K^{2}\tilde{p}$, which is the number of AR coefficients in a fully-parametrized VAR($\tilde{p}$) model.

In the first stage we execute group selection of AR coefficients by using PSC together with BIC. This use of group structure of AR coefficients effectively reduces the number of candidate models to be examined in the first stage. Similar use of the group structure of AR coefficients has also been employed in other settings, one of which is to determine the {\em Granger causality} between time series. This concept was first introduced by \citet{Granger1969} in econometrics.  It is shown that, e.g., see \citet{Helmut1991}, a Granger causal relationship can be examined by fitting VAR models to the multivariate time series in question, where non-zero AR coefficients indicate Granger causality between the corresponding series. In the literature, $l_{1}$-penalized regression (Lasso) has been widely used to explore sparsity in Granger causal relationships by shrinking AR coefficients to zero, e.g., see \citet{Arnold2008, Michailidis2010}. In particular, \citet{Lozano2009, Haufe2010} proposed to penalize groups of AR coefficients simultaneously, in which their use of the group structure of AR coefficients is similar to \eqref{groupARzero}. In spite of their common purpose of fitting sparse models, simulation results in Section \ref{section_simulation} will demonstrate the advantage of using PSC in conjunction with BIC over Lasso in discovering sparsity in AR coefficients. For detailed discussion on using VAR models to determine Granger causality, readers are referred to \citet{Granger1969, Helmut1991, Arnold2008}.

\subsection{Stage 2: refinement}
Stage 1 selects AR parameters related to the most conditionally correlated pairs of marginal series according to BIC. However, it may also have introduced spurious non-zero AR coefficients in the stage 1 model: As PSC can only be evaluated for pairs of series, we cannot select diagonal coefficients in $A_{1},\ldots, A_{p}$, nor can we select within the group of coefficients corresponding to one pair of component series. We therefore apply a second stage to further refine the stage 1 model. To eliminate these possibly spurious coefficients, the $(K+2\tilde{M})\tilde{p}$ non-zero AR coefficients of the stage 1 model are ranked according to the absolute values of their $t$-statistic. The $t$-statistic for a non-zero AR coefficient estimate $\hat{A}_{k}(i,j)$, ($k=1,\ldots,\tilde{p}$ and $i\neq j$) is,
\begin{equation}
t_{i,j,k}\defeq\frac{\hat{A}_{k}(i,j)}{\mathrm{s.e.}(\hat{A}_{k}(i,j))}. \label{tStat}
\end{equation}
Here the standard error of $\hat{A}_{k}(i,j)$ is computed from the asymptotic distribution of the constrained maximum likelihood estimator of the stage 1 model, which is, e.g., see \citet{Helmut1991},
\begin{equation}
\sqrt{T}(\hat{\alpha} - \alpha) \overset {d}{\Longrightarrow} N(0, ~\tilde{R}[\tilde{R}^{'}(\tilde{\Gamma}_{Y}(0)\otimes \tilde{\Sigma}_{Z}^{-1})\tilde{R}]^{-1}\tilde{R}^{'}), \label{sVARasympVar}
\end{equation}
where $\alpha\defeq\mathrm{vec}(A_{1},\ldots,A_{p})$ is the $K^2p\times 1$ vector obtained by column stacking the AR coefficient matrices $A_{1},\ldots,A_{p}$; $\hat{\alpha}$, $\tilde{\Gamma}_{Y}(0)$ and $\tilde{\Sigma}_{Z}$ are the maximum likelihood estimators of $\alpha$, $\Gamma_{Y}(0)\defeq\mathrm{cov}((Y_{t}^{'},\ldots,Y_{t-p+1}^{'})^{'})$ and $\Sigma_{Z}$, respectively; and $\tilde{R}$ is the {\em constraint matrix}, defined by equation \eqref{sVARequation} in Appendix \ref{sVARest}, of the stage 1 model. Therefore we can create a sequence $\mathbf{Q}_{2}$ of the $(K+2\tilde{M})\tilde{p}$ triplets $(i,j,k)$ by ranking the absolute values of the $t$-ratios \eqref{tStat} from highest to lowest. The AR coefficients corresponding to the {\em top} triplets in $\mathbf{Q}_{2}$ are more likely to be retained in the model because of their significance. In the second stage, there is only one parameter, denoted by $m$, controlling the complexity of the model, which is the number of non-zero AR coefficients to be retained.  And BIC is used to select the complexity of the final sVAR model. The steps of {\bf stage 2} are as follows.

Our 2-stage approach in the end leads to a sVAR model that contains $m^{*}$ non-zero AR coefficients corresponding to the top $m^{*}$ triplets in $\mathbf{Q}_{2}$. We denote this sVAR model by sVAR($p^{*}, m^{*}$), where $p^{*}$ is the order of autoregression and $m^{*}$ is the number of non-zero AR coefficients.
\begin{center}
\Ovalbox{\setlength{\itemsep}{0pt}
\begin{minipage}{0.97\textwidth}
\begin{center}
{\bf Stage 2}
\end{center}
\begin{itemize}\label{sVAR2stage}
\item[1.] Compute the $t$-statistic $t_{i,j,k}$ \eqref{tStat} for each of the $(K+2\tilde{M})\tilde{p}$ non-zero AR coefficient estimates of the stage 1 model.
\item[2.] Create a sequence $\mathbf{Q}_{2}$ of the $(K+2\tilde{M})\tilde{p}$ triplets $(i,j,k)$ by ranking $|t_{i,j,k}|$ from highest to lowest.
\item[3.] For each $m\in \{0,1,\ldots,(K+2\tilde{M})\tilde{p}\}$, consider the model that selects the $m$ non-zero AR coefficients corresponding to the top $m$ triplets in the sequence $\mathbf{Q}_{2}$. Under this parameter constraint, execute the constrained parameter estimation using results in Appendix \ref{sVARest} and compute the corresponding BIC according to $\mathrm{BIC}(m) = -2\log L + \log{T}\cdot m$.
\item[4.] Choose $m^{*}$ that gives the minimum BIC value. 
\end{itemize}
\end{minipage}}
\end{center}

\section{Numerical results}\label{section_results}
In this section, we provide numerical results on the performance of our 2-stage approach of fitting sVAR models. In Section \ref{section_simulation}, simulation results are presented to compare the performance of the 2-stage approach against competing Lasso-type methods of fitting sVAR models. In Section \ref{section_real}, the 2-stage approach is applied to two real data examples. The first is the Google Flu Trends data and the second is a time series of concentration levels of air pollutants.

\subsection{Simulation}\label{section_simulation}
Simulation results are presented to demonstrate the performance of our 2-stage approach of fitting sVAR models.
We compare the 2-stage approach with Lasso-VAR methods. To apply Lasso-VAR methods, the VAR model is re-formulated as a linear regression problem, where current values of the time series are treated as the response variable and lagged values are treated as the explanatory variables. Then Lasso can be applied to select the AR coefficients and fit sVAR models, e.g., see \citet{Sosa2005, Hsu2008, Arnold2008, Lozano2009, Haufe2010, Michailidis2010, Song2011}. The Lasso method shrinks the AR coefficients towards zero by minimizing a target function, which is the sum of a loss function and a $l_{1}$ penalty on the AR coefficients. Unlike linear regression models, the choice of the loss function between the sum of squared residuals and the minus log likelihood will affect the resulted Lasso-VAR models even if the multivariate time series is Gaussian. This is because the noise covariance matrix $\Sigma_{Z}$ is taken into account in the likelihood function of a Gaussian VAR process but not in the sum of squared residuals. In general, this distinction will lead to different VAR models unless the unknown covariance matrix $\Sigma_{Z}$ equals to a scalar multiple of the identity matrix, e.g., see Appendix \ref{lassoToVAR}. We notice that this issue of choosing the loss function has not been addressed in the literature of Lasso-VAR models. For example, \citet{Arnold2008, Lozano2009, Haufe2010, Michailidis2010, Song2011} all used the sum of squared residuals as the loss function and did not consider the possibility of choosing the minus log likelihood as the loss function. The simulation setups in these papers all assume, either explicitly or implicitly, that the covariance matrix $\Sigma_{Z}$ is diagonal or simply the identity matrix. Therefore in our simulation we apply Lasso to VAR modeling under both cases: in the first case we choose the sum of squared residuals as the loss function and denote it as the Lasso-SS method; in the second case we use the minus log likelihood as the loss function and denote it as the Lasso-LL method. Details of fitting these two Lasso-VAR models are given in Appendix \ref{lassoToVAR}.

The Lasso-VAR approach simultaneously performs model selection and parameter estimation, which is usually considered as an advantage of the approach. However, our simulation results suggest that simultaneous model selection and parameter estimation can weaken the performance of the Lasso-VAR approach. This is because Lasso-VAR methods, such as Lasso-SS and Lasso-LL, have a tendency to over-select the autoregression order of VAR models, a phenomenon reported by many, see \citet{Arnold2008, Lozano2009, Michailidis2010}. This over-specified model complexity potentially increases the mean squared error of the AR coefficient estimates of Lasso-VAR models. On the contrary, simulation results show that our 2-stage approach is able to identify the correct set of non-zero AR coefficients more often and it also achieves better parameter estimation efficiency than the two competing Lasso-VAR methods. In addition, simulation results also suggest that the Lasso-SS method, which does not take into account the noise covariance matrix $\Sigma_{Z}$ in its model fitting, performs the worst among the three.

Here we describe the simulation example used to compare the performance of our 2-stage approach, the Lasso-SS and the Lasso-LL methods of fitting sVAR models. Consider the $6$-dimensional VAR(1) process $\{Y_{t}\}=\{(Y_{t,1},\ldots,Y_{t,6})^{'}\}$ given by,
\small
\begin{equation}\label{sim1equation}
\left(
\begin{array}{c}
Y_{t,1}\\
Y_{t,2}\\
Y_{t,3}\\
Y_{t,4}\\
Y_{t,5}\\
Y_{t,6}\\
\end{array}
\right)
 =
\left(
\begin{array}{cccccc}
0.8 & 0  & 0  & 0    & 0   & 0\\
0   & 0   & 0  & 0.3 & 0   & 0\\
0   & 0   & 0  &0    &-0.3 & 0\\
0.6 &0   & 0  &0    & 0    & 0\\
0   & 0  &0.6 & 0   & 0    & 0\\
0   & 0  &0   & 0   & 0    &  0.8\\
\end{array}
\right)
\left(
\begin{array}{c}
Y_{t-1,1}\\
Y_{t-1,2}\\
Y_{t-1,3}\\
Y_{t-1,4}\\
Y_{t-1,5}\\
Y_{t-1,6}\\
\end{array}
\right)
+
\left(
\begin{array}{c}
Z_{t,1}\\
Z_{t,2}\\
Z_{t,3}\\
Z_{t,4}\\
Z_{t,5}\\
Z_{t,6}\\
\end{array}
\right),
\end{equation}
\normalsize
where $Z_{t}=(Z_{t,1},\ldots,Z_{t,6})^{'}$ are iid Gaussian noise with mean {\bf 0} and covariance matrix $\Sigma_{Z}$. The order of autoregression in \eqref{sim1equation} is $p=1$ and there are 6 non-zero AR coefficients, so \eqref{sim1equation} specifies a sVAR$(1,6)$ model. The covariance matrix $\Sigma_{Z}$ of the Gaussian noise is,
\begin{equation*}
\Sigma_{Z} =
\left(
\begin{array}{cccccc}
\delta^2& \delta/4  & \delta/6 & \delta/8 & \delta/10 & \delta/12 \\
\delta/4  & 1   & 0 & 0 & 0 & 0\\
\delta/6  & 0  & 1  &0&0 &0\\
\delta/8  &0  &0   &1& 0 &0\\
\delta/10  & 0  &0& 0 &1 & 0\\
\delta/12 & 0  &0 & 0 &0 & 1\\
\end{array}
\right).
\end{equation*}
We can see that the marginal series $\{Y_{t,1}\}$ is related to all other series via $\Sigma_{Z}$. And we can change the value of $\delta^2$ to compare the impact of the variability of $\{Y_{t,1}\}$ on the performance of the three competing methods. We compare the three methods according to five metrics: (1) the selected order of autoregression $\hat{p}$; (2) the number of non-zero AR coefficient estimates $\hat{m}$; (3) the squared bias of the AR coefficient estimates,
\begin{center}
$\displaystyle\sum_{k=1}^{p\vee\hat{p}}\sum_{i,j=1}^{K}[\mathbf{E}[\hat{A}_{k}(i,j)] - A_{k}(i,j)]^2$;
\end{center}
(4) the variance of the AR coefficient estimates,
\begin{center}
$\displaystyle\sum_{k=1}^{p\vee\hat{p}}\sum_{i,j=1}^{K}\mathrm{var}(\hat{A}_{k}(i,j))$;
\end{center} and (5) the mean squared error (MSE) of the AR coefficient estimates,
\begin{center}
$\displaystyle\sum_{k=1}^{p\vee\hat{p}}\sum_{i,j=1}^{K}\{[\mathbf{E}[\hat{A}_{k}(i,j)] - A_{k}(i,j)]^2 + \mathrm{var}(\hat{A}_{k}(i,j))\}$,
\end{center}
where $p\vee\hat{p}\defeq\max\{p,\hat{p}\}$ and $A_{k}(i,j)\defeq0$ for any triplet $(k,i,j)$ such that $k>1$ and $1\le i,j \le K$. The first two metrics show the model selection performance and the latter three metrics reflect the efficiency of parameter estimates of each method. The pre-specified range of the autoregression order $p$ is $\mathbb{P}=\{0,1,2,3\}$. Selection of the tuning parameter for the two Lasso-VAR methods is based on ten-fold cross validations, as described in Appendix \ref{lassoToVAR}. We let $\delta^2$ in $\Sigma_{Z}$ take values from $\{1,4,25,100\}$. The sample size $T$ is 100 and results are based on 500 replications.

The five metrics for comparison are summarized in Table \ref{sim1_fiveQuantTable}. The $\hat{p}$ column shows that the 2-stage approach is able to correctly select the autoregression order $p=1$ while the two Lasso-VAR methods over-select the autoregression order. Furthermore, the true number of non-zero AR coefficients is $m=6$. As shown by the $\hat{m}$ column, the average number of non-zero AR coefficient estimates from the 2-stage approach is very close to 6. At the same time, this number from either the Lasso-SS or the Lasso-LL method is much larger than 6, meaning that the two Lasso-VAR methods lead to a lot of spurious non-zero AR coefficients. Second, we compare the efficiency of parameter estimates. The $\mathrm{bias}^2$ column shows that the 2-stage approach has much smaller estimation bias than the two Lasso-VAR methods. This is because the $l_{1}$ penalty is known to produce large estimation bias for large non-zero coefficients, see \citet{Fan2001}. In addition, the large number of spurious non-zero AR coefficients also increases the variability of the parameter estimates from the two Lasso-VAR methods. This is reflected in the $\mathrm{variance}$ column, showing that the variance of the AR coefficient estimates from the Lasso-SS and the Lasso-LL methods are larger than that from the 2-stage approach. Therefore the 2-stage approach has a much smaller MSE than the two Lasso-VAR methods. And this difference in MSE becomes more notable as the marginal variability $\delta^2$ increases.

\begin{table}
\begin{center}
\begin{tabular}{l|r|r|r|r|r|r}
\multicolumn{2}{c|}{~} & \quad $\hat{p}$\quad\quad & \quad$\hat{m}\quad$\quad &\quad bias$^2$\quad & variance \quad& \quad MSE \quad \\
\hline

\cline{2-7}
                           & 2-stage & 1.000  &  5.854  &  0.021  & 0.092 & 0.113  \\
\cline{2-7}
$\delta^2=1$        & Lasso-LL   & 1.208 &  17.852  &  0.060  &   0.099  &  0.159  \\
\cline{2-7}
                            & Lasso-SS   & 1.218    & 17.156   &  0.054   &  0.092  & 0.146 \\
\hline
\cline{2-7}
                            & 2-stage &  1.000 &  6.198  & 0.006  &  0.087 & 0.093   \\
\cline{2-7}
$\delta^2=4$        & Lasso-LL  & 1.150 &  17.254 &  0.046  &  0.103  &  0.149 \\
\cline{2-7}
                            & Lasso-SS  & 1.246   & 16.478  & 0.053   & 0.136   & 0.188 \\
\hline
\cline{2-7}
                           & 2-stage & 1.000  & 6.190   & 0.002  &  0.073 &  0.075  \\
\cline{2-7}
$\delta^2=25$      & Lasso-LL  &  1.179  & 17.275 & 0.042  & 0.274  &   0.316 \\
\cline{2-7}
				      & Lasso-SS  & 1.364   & 14.836  & 0.094   & 0.875 &  0.969  \\
\hline
\cline{2-7}
                            & 2-stage & 1.000  & 6.260   & 0.003  & 0.175 &  0.178 \\
\cline{2-7}
$\delta^2=100$     & Lasso-LL   &  1.203  & 17.464  & 0.056  & 0.769 &  0.825  \\
\cline{2-7}
                            & Lasso-SS    & 1.392  & 11.108  &  0.298   & 2.402  & 2.700 \\
\hline
\end{tabular}
\caption{The five metrics from the 2-stage approach, the Lasso-LL and the Lasso-SS methods. (1) $\hat{p}$: the average selected autoregression order. (2) $\hat{m}$: the average number of non-zero AR coefficient estimates. (3) bias$^{2}$, (4) variance, (5) MSE: the squared bias, the variance and the MSE of the AR coefficient estimates, respectively.}
\label{sim1_fiveQuantTable}
\end{center}
\end{table}

\begin{figure}[bt]
\begin{center}
$\begin{array}{ccc}
\subfloat[AR coefficients]{\includegraphics[width=0.31\textwidth]{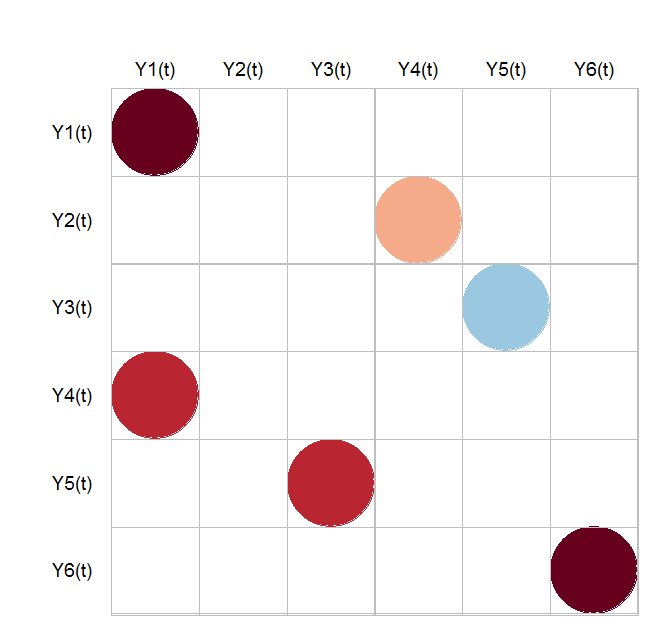}} &
\subfloat[stage 1 ($\delta^2=1$)]{\includegraphics[width=0.31\textwidth]{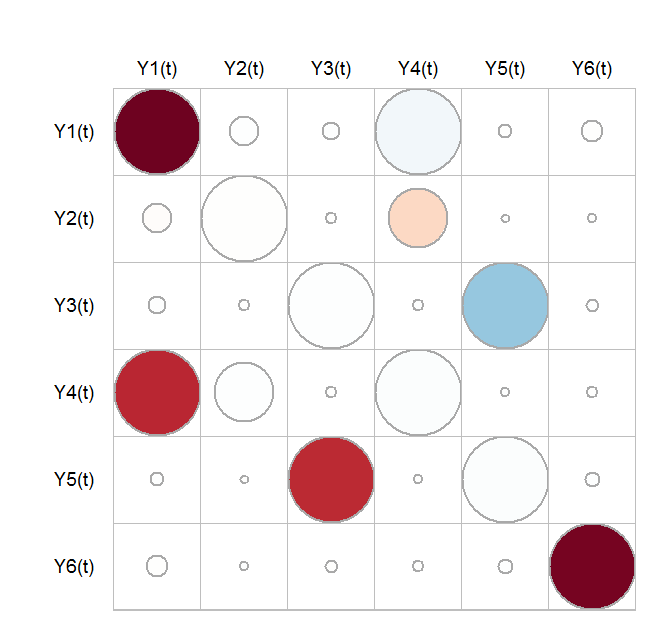}}  &
\subfloat[stage 2 ($\delta^2=1$)]{\includegraphics[width=0.31\textwidth]{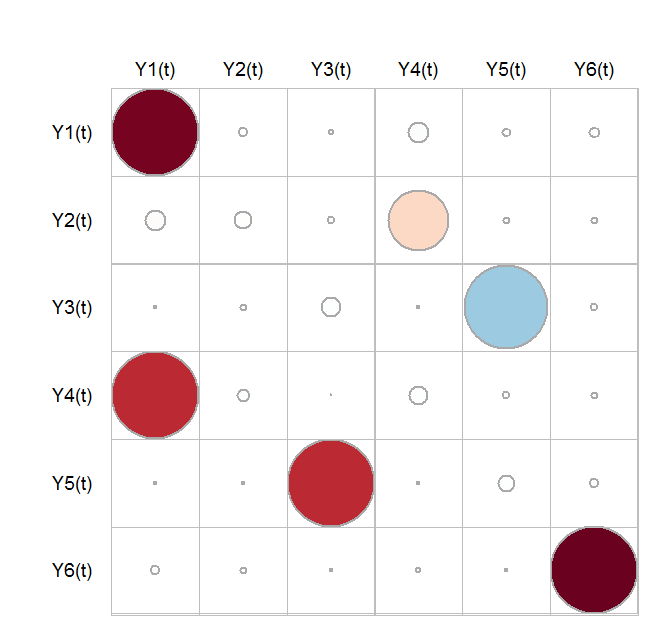}}   \\
\subfloat{\includegraphics[width=0.31\textwidth]{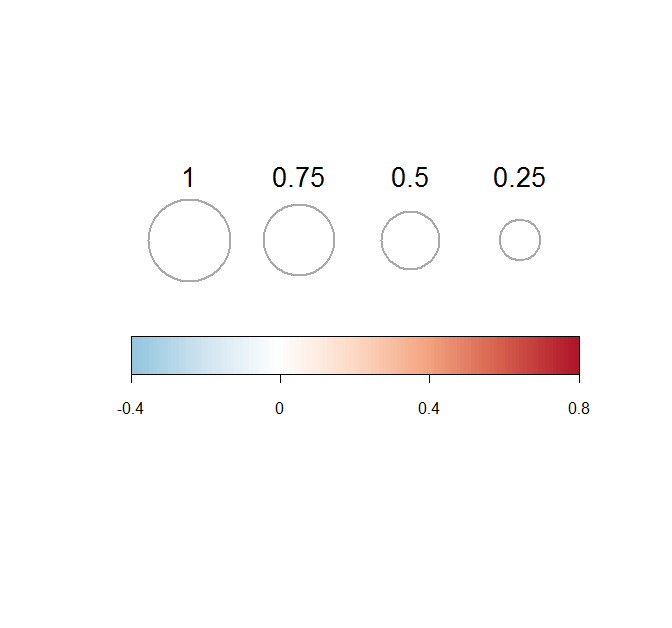}} &
\subfloat[Lasso-LL ($\delta^2=1$)]{\includegraphics[width=0.31\textwidth]{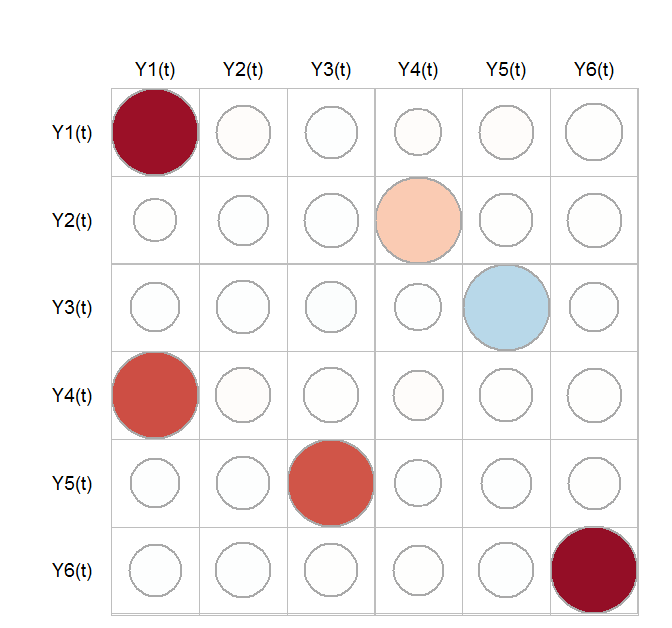}}  &
\subfloat[Lasso-SS ($\delta^2=1$)]{\includegraphics[width=0.31\textwidth]{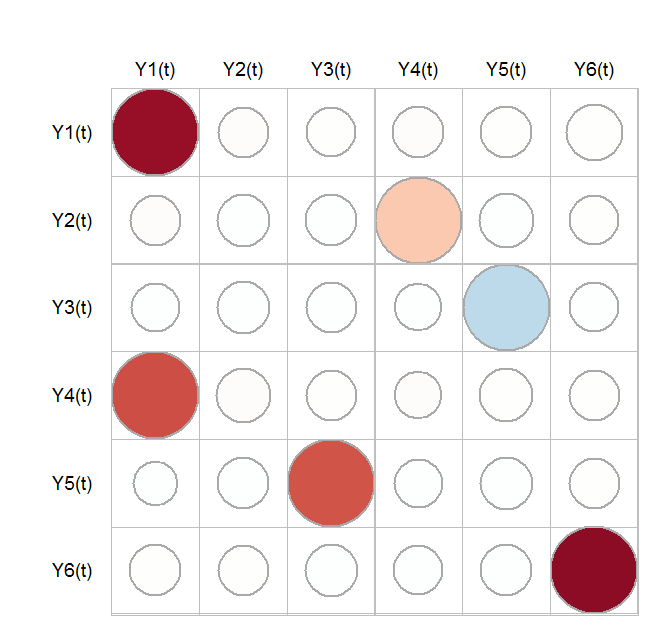}}
\end{array}$
\caption{Displays of the AR coefficient estimates from stages 1 and 2 of the 2-stage approach, the Lasso-LL and the Lasso-SS methods when $\delta^2=1$. Panel (a) displays the true AR coefficient matrix $A_{1}$, where the color of each circle shows the true value of the corresponding AR coefficient. In panels (b), (c), (e) and (f), the size of each circle is proportional to the percent of times (out of 500 replications) the corresponding AR coefficient is selected; the color of each circle shows the average of the 500 estimates of that AR coefficient.}
\label{sim1_MatrixGraph_true_stage1_stage2_LassoSS_LassoLL_delta1}
\end{center}
\end{figure}

A comparison of the AR coefficient estimation performance when $\delta^2=1$ is displayed in Figure \ref{sim1_MatrixGraph_true_stage1_stage2_LassoSS_LassoLL_delta1}. Panels (b) and (c) of Figure \ref{sim1_MatrixGraph_true_stage1_stage2_LassoSS_LassoLL_delta1}
show the AR coefficient estimates from stages 1 and 2 of the 2-stage approach. The size of each circle is proportional to the percent of times (out of 500 replications) the corresponding AR coefficient is selected and the color of each circle shows the average of the 500 estimates of that AR coefficient. For comparison, panel (a) displays the true AR coefficient matrix $A_{1}$, where the color of a circle shows the true value of the corresponding AR coefficient. We can see from panel (b) that the first stage is able to select the AR coefficients belonging to pairs of conditionally correlated marginal series. But the stage 1 model contains spurious non-zero AR coefficients, as indicated by the presence of 6 dominant white circles in panel (b) at 4 diagonal positions, i.e., $(2,2), (3,3) ,(4,4), (5,5)$, and 2 off-diagonal positions, i.e., $(1,4), (4,2)$. These white circles effectively disappear in panel (c) due to the second stage refinement. This observation demonstrates the effectiveness of the second stage refinement. In addition, the similarity between panel (a) and panel (c) has two implications: first, the presence of 6 dominant color circles in both panels suggests that the 2-stage approach is able to select the true non-zero AR coefficients with high probabilities; second, the other tiny circles in panel (c) indicate that the 2-stage approach leads to only a small number of spurious AR coefficients. These two implications together show that the 2-stage approach is able to correctly select the non-zero AR coefficients for this sVAR model. On the other hand, panels (e) and (f) display the estimated AR coefficients from the Lasso-LL and the Lasso-SS methods, respectively. The most notable aspect in these two panels is the prevalence of medium-sized white circles. The whiteness of these circles indicates that the corresponding AR coefficient estimates are unbiased. However, according to the legend panel, the size of these circles corresponds to an approximate 50$\%$ chance that each of these truly zero AR coefficients is selected by the Lasso-VAR methods. As a result, both two Lasso-VAR methods lead to a large number of spurious non-zero AR coefficients and their model selection results are highly variable. Consequently, it is more difficult to interpret these Lasso-VAR models. This observed tendency for Lasso-VAR methods to over-select the non-zero AR coefficients is consistent with the numerical findings in \citet{Arnold2008, Lozano2009, Michailidis2010}.

\begin{figure}[bt]
\begin{center}$
\begin{array}{ccc}
\subfloat[2-stage ($\delta^2=4$)]{\includegraphics[width=0.3\textwidth]{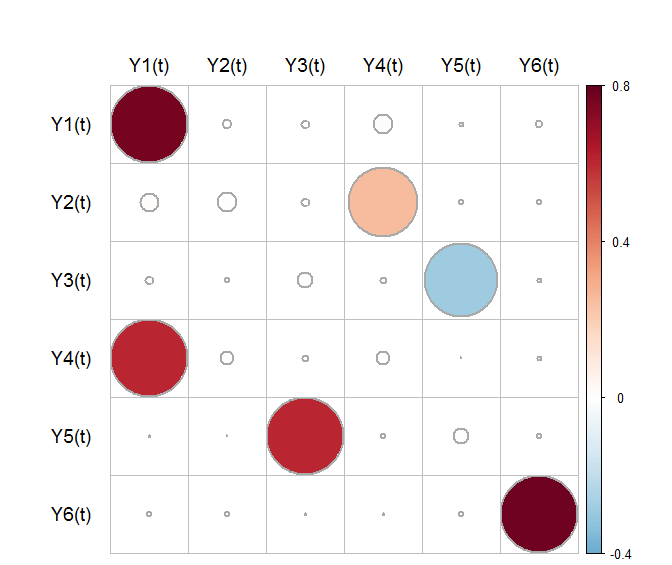}}  &
\subfloat[2-stage ($\delta^2=25$)]{\includegraphics[width=0.3\textwidth]{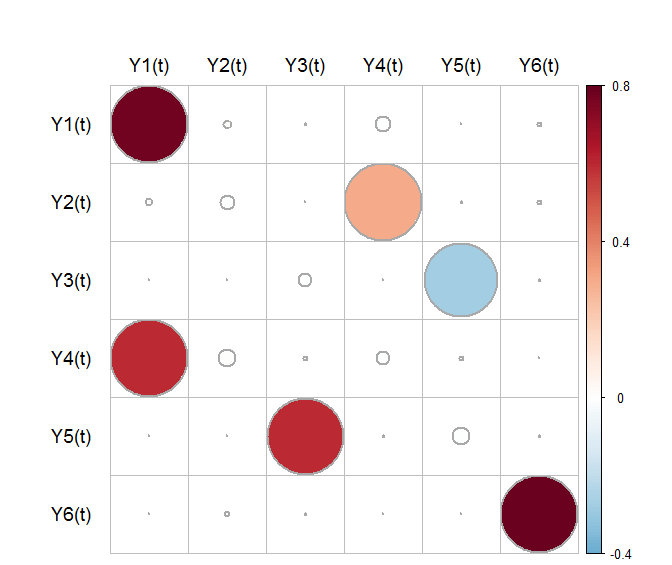}}  &
\subfloat[2-stage ($\delta^2=100$)]{\includegraphics[width=0.3\textwidth]{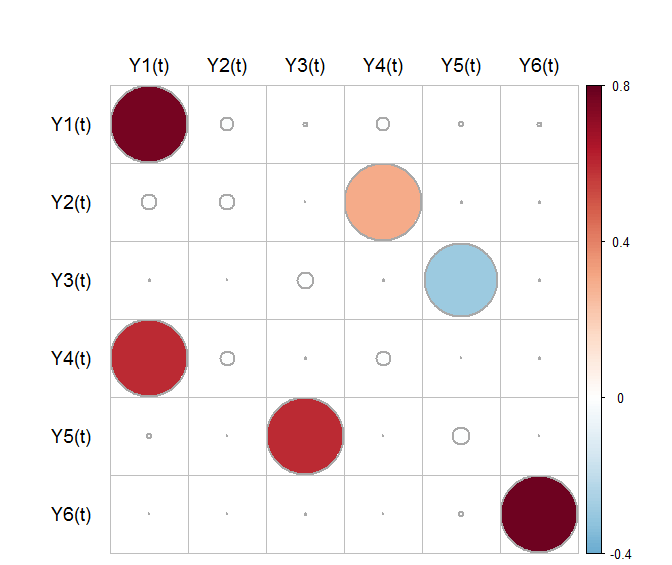}}  \\
\subfloat[Lasso-LL ($\delta^2=4$)]{\includegraphics[width=0.3\textwidth]{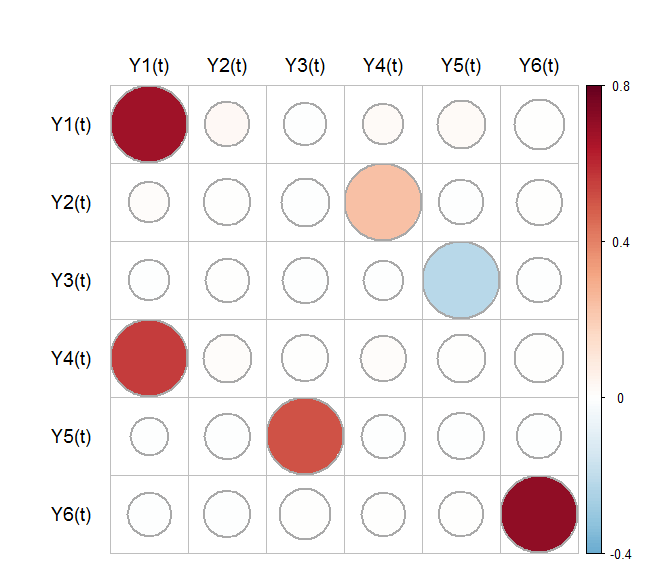}}  &
\subfloat[Lasso-LL ($\delta^2=25$)]{\includegraphics[width=0.3\textwidth]{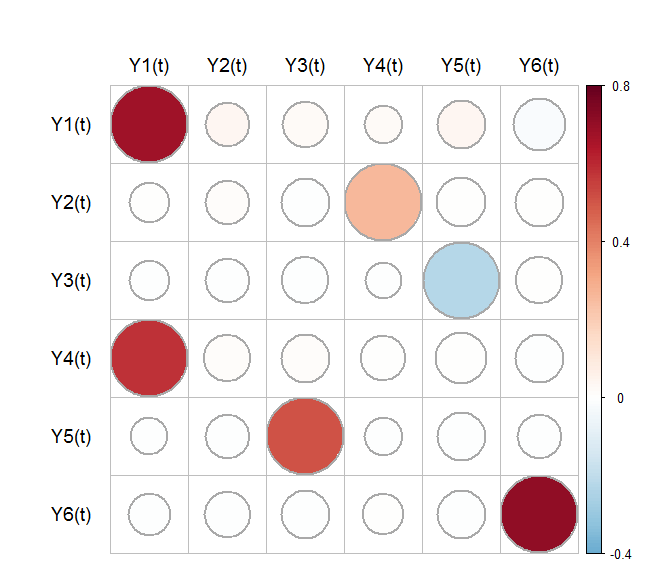}}  &
\subfloat[Lasso-LL ($\delta^2=100$)]{\includegraphics[width=0.3\textwidth]{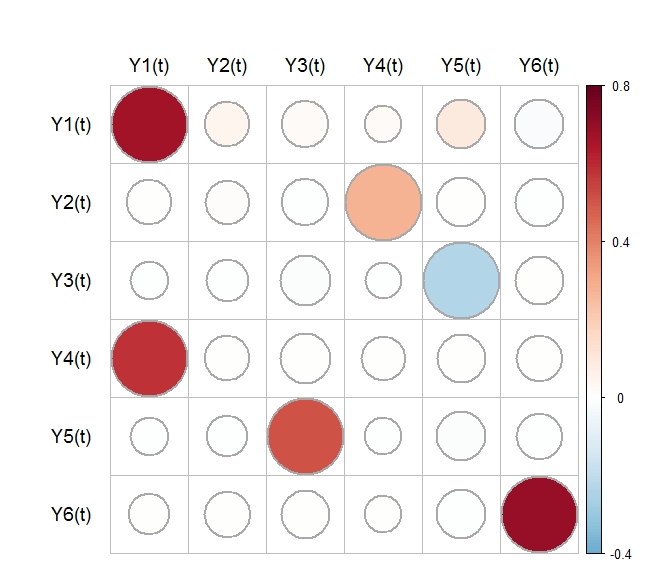}} \\
\subfloat[Lasso-SS ($\delta^2=4$)]{\includegraphics[width=0.3\textwidth]{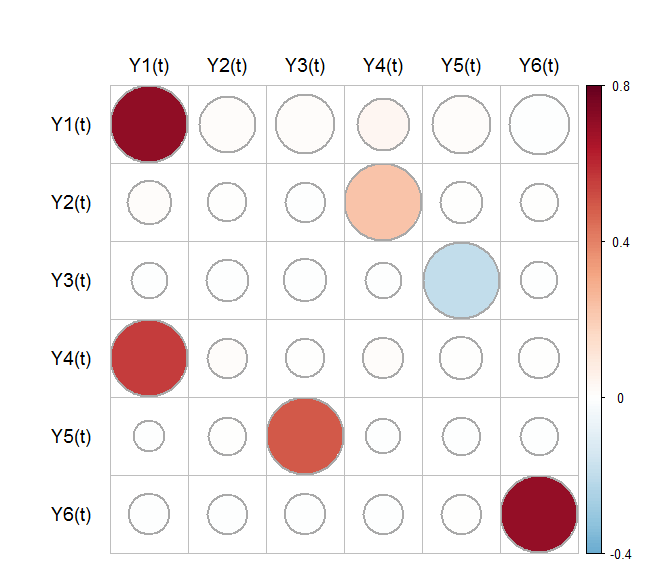}}  &
\subfloat[Lasso-SS ($\delta^2=25$)]{\includegraphics[width=0.3\textwidth]{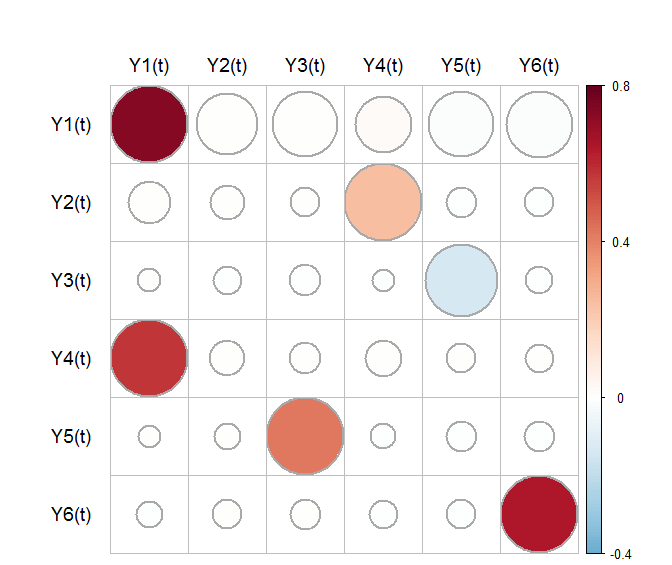}} &
\subfloat[Lasso-SS ($\delta^2=100$)]{\includegraphics[width=0.3\textwidth]{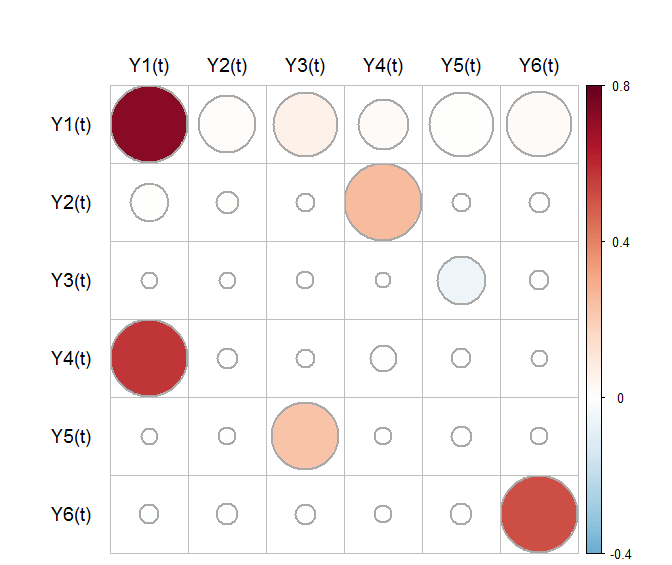}}
\end{array}$
\caption{Displays of the AR coefficient estimates from stages 1 and 2 of the 2-stage approach, the Lasso-LL and the Lasso-SS methods when $\delta^2=4,25$ and $100$, respectively. The interpretation of the size and the color of a circle is the same as in Figure \ref{sim1_MatrixGraph_true_stage1_stage2_LassoSS_LassoLL_delta1}.}
\label{sim1_MatrixGraph_2stage_LassoLL_LassoSS_delta4_delta25_delta100}
\end{center}
\end{figure}

We also compare the impact of the marginal variability of $\{Y_{1,t}\}$ on the performance of each method. Figure \ref{sim1_MatrixGraph_2stage_LassoLL_LassoSS_delta4_delta25_delta100} displays the estimated AR coefficients from the 2-stage approach as well as the two Lasso-type methods for $\delta^2=4, 25$ and $100$, respectively. We can see that the performance of the 2-stage approach remains persistently good against the changing marginal variability $\delta^2$. This is because the 2-stage approach involves estimating the covariance matrix $\Sigma_{Z}$ and therefore will adjust for the changing variability. On the other hand, both Lasso-VAR methods persistently over-select the AR coefficients as $\delta^2$ varies. But it is interesting to notice that the impact of the changing variability is different for the Lasso-SS and the Lasso-LL methods. The model selection result of the Lasso-SS method is severely impacted by the changing variability. From panels (g), (h) and (i), we can see that as $\delta^2$ increases from 4 to 100, the size of the white circles in the first row increases while the size of the white circles in the other five rows decreases. This observation suggests that as the marginal variability of $\{Y_{t,1}\}$ increases, the Lasso-SS method will increasingly over-estimate the temporal influence of the other 5 marginal series into $\{Y_{t,1}\}$ and leads to spurious AR coefficients in the first row of $A_{1}$. On the other hand, panels (d), (e) and (f) show that the model selection result of the Lasso-LL method is not much influenced by the changing variability. Such a difference between the Lasso-SS and the Lasso-LL methods is due to the fact that the Lasso-LL method takes into account the covariance matrix $\Sigma_{Z}$
while the Lasso-SS method does not. The observed distinction between the Lasso-SS and the Lasso-LL methods verifies that the choice of the loss function will affect the resulted Lasso-VAR model, a fact that has not been addressed in the literature of Lasso-VAR modeling. In this simulation example, the Lasso-LL method benefits from modeling the covariance matrix $\Sigma_{Z}$ and is superior to the Lasso-SS method.

\begin{figure}[bt]
\begin{center}
\includegraphics[width=0.9\textwidth,height=0.7\textwidth]{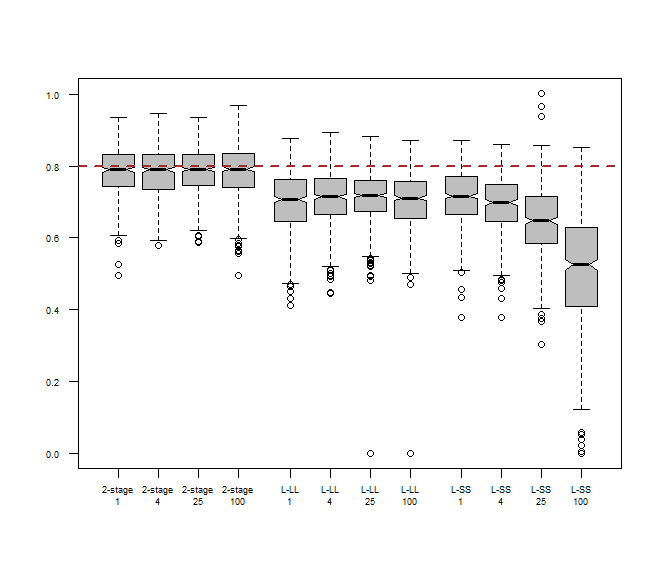}
\caption{Sampling distributions of the estimators of $A_{1}(6,6)$ from the 2-stage approach (the left 4 boxplots), the Lasso-LL method (the middle 4 boxplots) and the Lasso-SS method (the right 4 boxplots) for $\delta^2=1, 4, 25$ and $100$, respectively. The dashed horizontal line indicates the true value of $A_{1}(6,6)=0.8$.}
\label{sim1_parEst_A1_66_2stage_LassoLL_LassoSS_boxplot}
\end{center}
\end{figure}

Finally, we investigate the estimators of one particular AR coefficient from the three methods in more detail.
Figure \ref{sim1_parEst_A1_66_2stage_LassoLL_LassoSS_boxplot} displays the sampling distributions of the estimator $\hat{A}_{1}(6,6)$ from the 2-stage approach as well as the two Lasso-VAR methods for $\delta^2=1, 4, 25$ and $100$, respectively. Estimation of $A_{1}(6,6)$ is of interest because the marginal series $\{Y_{t,6}\}$ is exclusively driven by its own past values. Ideally, due to such ``isolation", the estimation of $A_{1}(6,6)$ should not be impacted much by the estimation of the AR coefficients in the $5\times 5$ upper-left sub-matrix of $A_{1}$. Moreover, $A_{1}(6,6)$ has a large true value of 0.8 and it is interesting to compare the estimation bias for this large AR coefficient. Figure \ref{sim1_parEst_A1_66_2stage_LassoLL_LassoSS_boxplot} shows that the estimators of $A_{1}(6,6)$ from the 2-stage approach and the Lasso-LL method are not impacted much by the changing variability of $\{Y_{t,1}\}$. But the Lasso-SS estimator for $A_{1}(6,6)$ becomes more biased and volatile as the marginal variability increases from $\delta^2=1$ to $\delta^2=100$. Although both the 2-stage sVAR and the Lasso-LL estimators of $A_{1}(6,6)$ are robust to the changing values of $\delta^2$, the difference between their bias is significant. The 2-stage approach gives an estimator of $A_{1}(6,6)$ that remains nearly unbiased as $\delta^2$ varies. However, there is a systematic bias in the Lasso-LL estimator of $A_{1}(6,6)$, which is due to the shrinkage effect of the Lasso penalty on the selected AR coefficients.

\subsection{Real data examples}\label{section_real}
{\bf Google Flu Trends data}. In this example, we consider the {\em Google Flu Trends} data, which can be viewed as a measure of the level of influenza activity in the US. It has been noticed by many researchers that the frequencies of certain Internet search terms can be predictive of the influenza activity within a future time period, e.g., see \citet{Polgreen2008, Eysenbach2009, Hulth2009}. Based on this fact, a group of researchers at Google applied logistic regression to select the top 45 Google user search terms that are most indicative of the influenza activity. These selected 45 terms were then used to produce the Google Flu Trends data, see \citet{Ginsberg2009}. The Google Flu Trends data consist of weekly predicted numbers of influenza-like-illness (ILI) \footnote{According to the Centers for Disease Control and Surveillance, an influenza-like-illness is defined as a fever of 100 degrees Fahrenheit (or higher) along with a cough and/or sore throat in the absence of a known cause other than influenza.} related visits out of every 100,000 random outpatient visits within a US region. The Google Flu Trends prediction has been shown to be highly consistent with the ILI rate reported by the Centers for Disease Control and Surveillance (CDC), where the ILI rate is the probability that a random outpatient visit is related to an influenza-like-illness. But the Google Flu Trends data have two advantages over the traditional CDC influenza surveillance report: first, the Google Flu Trends predictions are available 1 or 2 weeks before the CDC report is published and therefore provide a possibility to forecast the potential outbreak of influenza epidemics; second, since Google is able to map the IP address of each Google user search to a specific geographic area, the Google Flu Trends data enjoy a finer geographic resolution than the CDC report. In particular, the Google Flu Trends data are published not only at the US national level but are also available for the 50 states, the District of Columbia and 122 cities throughout the US. In contrast, the CDC surveillance report is available only at the national level and for 10 major US regions (each region is a group of states). Due to these advantages, there has been increasing interest in modeling the Google Flu Trends data to help monitor the influenza activity in the US, e.g., see \citet{Lopes2010, Fox2011}.

\begin{figure}[!h]
\begin{center}$
\begin{array}{cc}
\subfloat[BIC curve of stage 1]{\includegraphics[width=0.48\textwidth]{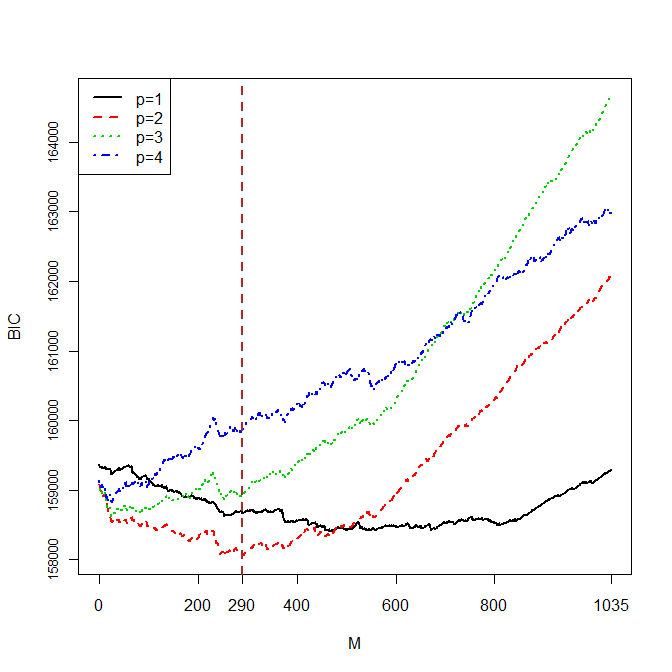}}  &
\subfloat[BIC curve of stage 2]{\includegraphics[width=0.48\textwidth]{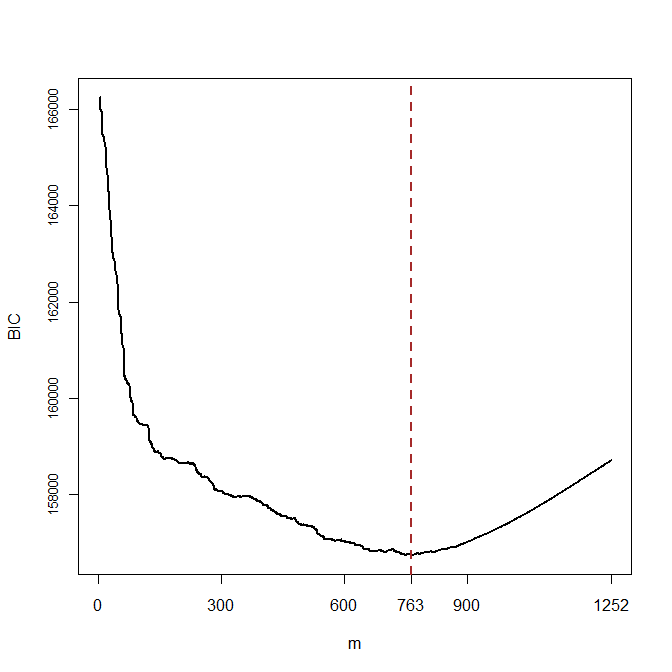}}
\end{array}$
\caption{BIC curves of stages 1 and 2 of the 2-stage approach. In panel (a), the x-axis $M$ refers to the number of top pairs selected. Each curve corresponds to one autoregression order $p \in \{1,2,3,4\}$ and shows the BIC values as $M$ varies from $0$ to $1035=\binom{46}{2}$. The BIC value of $p=0$ is not shown since it is much higher. In panel (b), the x-axis $m$ refers to the number of non-zero AR coefficients retained and the curve shows the BIC values as $m$ varies from 0 to 1252. In both panels, the dashed vertical line indicates where the minimum BIC value occurs.}
\label{googleFluTrends_2006to2010_BIC_stage1_stage2}
\end{center}
\end{figure}

We apply the 2-stage approach to fit a sVAR model to the weekly Goolge Flu Trends data from the week of January 1, 2006 to the week of December 26, 2010, so the sample size is $T=260$. Out of the 51 regions (50 states and the District of Columbia), we remove 5 states (Alaska, Hawaii, North Dakota, South Dakota and Wyoming) from our analysis due to incompleteness of the data during the selected time period. So the dimension of the process in this example is $K=46$ and we refer to these 46 regions as 46 states for simplicity. In applying the 2-stage approach, the pre-specified range of the autoregression order $p$ is $\mathbb{P}=\{0,1,2,3,4\}$. The 2-stage approach leads to a sVAR(2,763) model, which has only as many as $19.30\%=763/(46^2\times2)$ of the AR coefficients in a fully-parametrized VAR(2) model. Figure \ref{googleFluTrends_2006to2010_BIC_stage1_stage2} displays the BIC curves from stages 1 and 2 of the 2-stage approach, respectively. From panel (a) of stage 1, we can see that the first stage selects the autoregression order $\tilde{p}=2$ and $\tilde{M}=290$ pairs of distinct marginal series into the model. So the stage 1 model contains $(K+2\tilde{M})\tilde{p}=(46+290\cdot 2)\cdot 2=1252$ non-zero AR coefficients. The second stage follows by further selecting $m^{*}=763$ non-zero AR coefficients and leads to the final sVAR(2,763) model. For comparison, we also fit an unrestricted VAR(2) model and apply the Lasso-SS method to fit another sVAR model. 
Based on a ten-fold cross validation, the Lasso-SS method results in a VAR model with 3123 non-zero AR coefficients, which we denote as Lasso-SS(2,3123).

\begin{figure}[p]
\begin{center}$
\begin{array}{cc}
\subfloat[$\hat{A}_{1}$ in VAR(2)]{\includegraphics[width=0.45\textwidth]{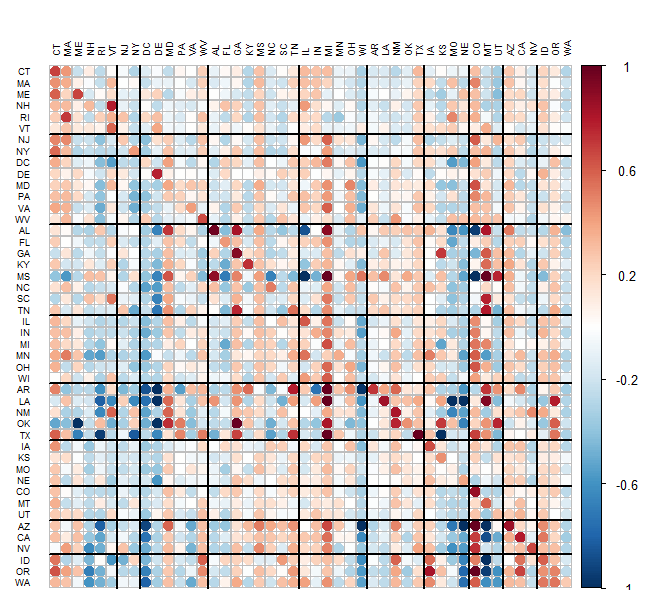}}  &
\subfloat[$\hat{A}_{2}$ in VAR(2)]{\includegraphics[width=0.45\textwidth]{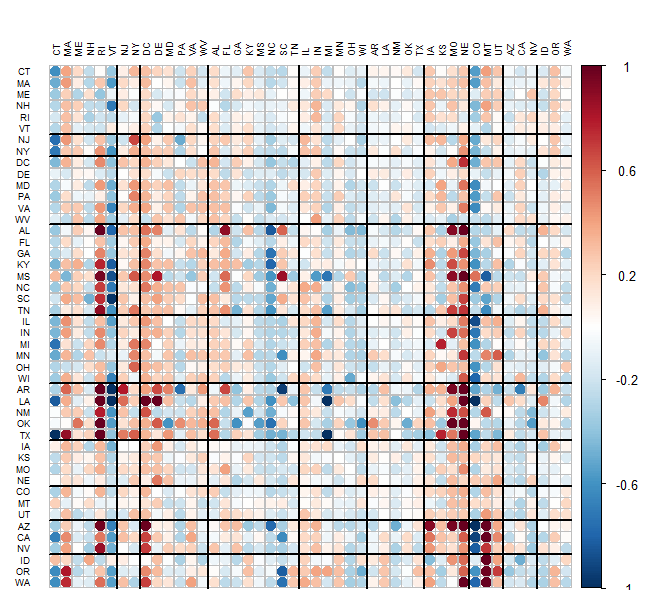}} \\
\subfloat[$\hat{A}_{1}$ in sVAR(2,763)]{\includegraphics[width=0.45\textwidth]{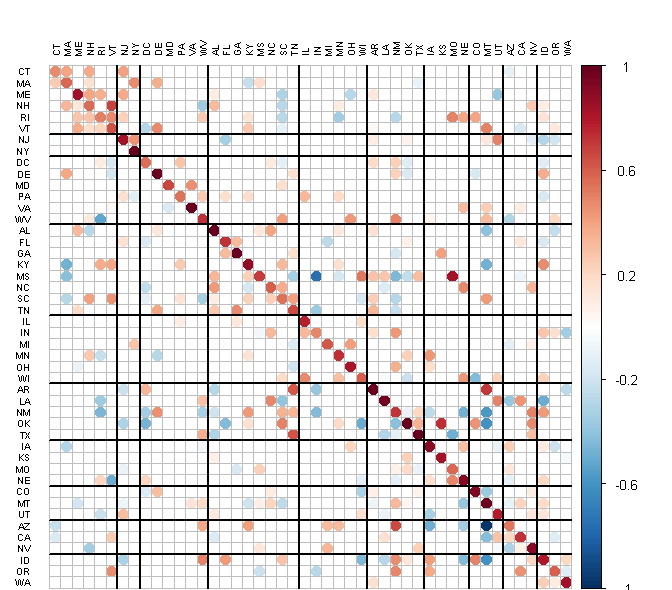}}  &
\subfloat[$\hat{A}_{2}$ in sVAR(2,763)]{\includegraphics[width=0.45\textwidth]{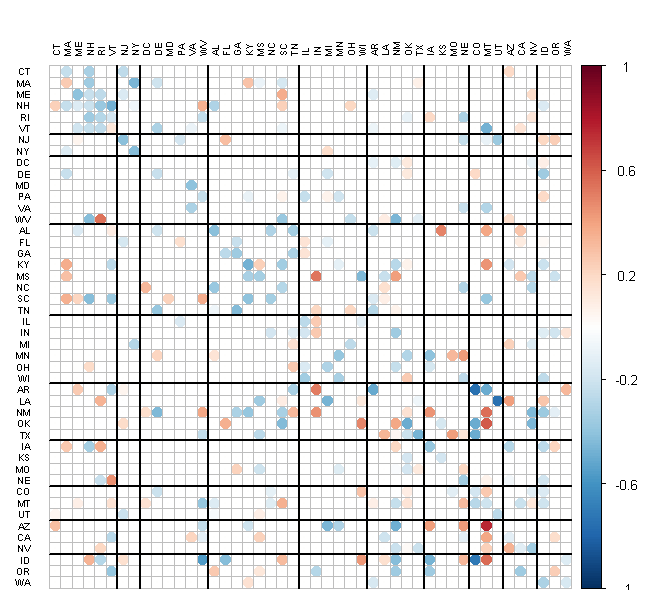}}  \\
\subfloat[$\hat{A}_{1}$ in Lasso-SS(2,3123)]{\includegraphics[width=0.45\textwidth]{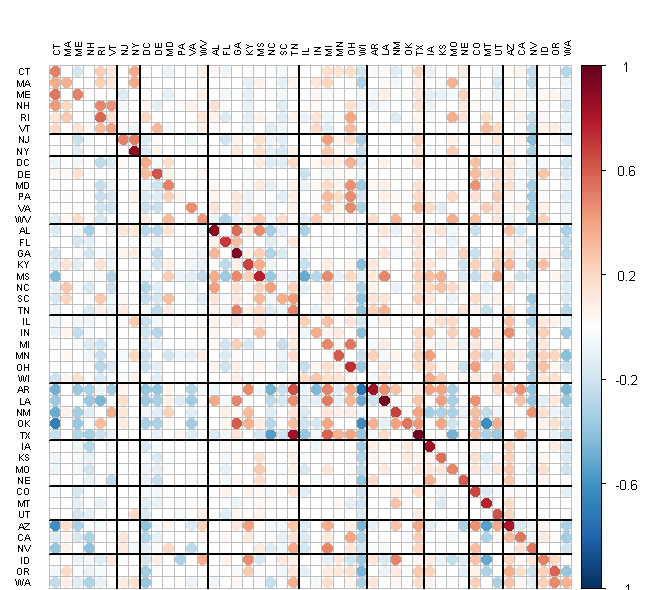}}  &
\subfloat[$\hat{A}_{2}$ in Lasso-SS(2,3123)]{\includegraphics[width=0.45\textwidth]{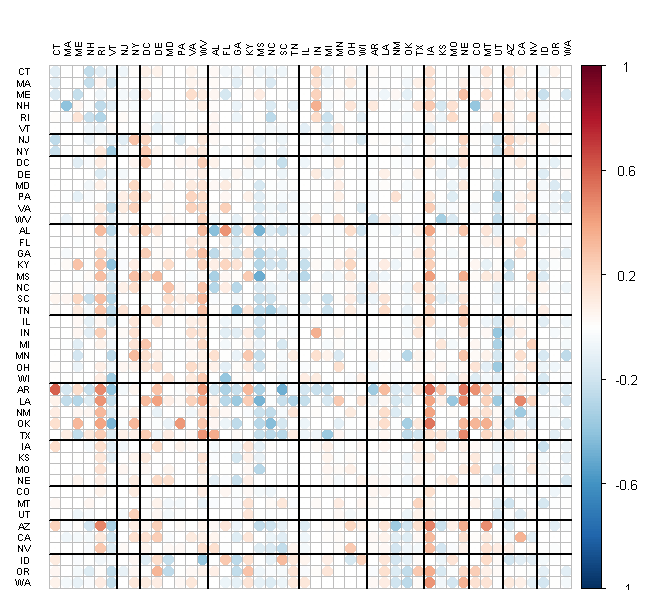}}  \\
\end{array}$
\caption{Displays of the AR coefficient estimates from the VAR(2), the sVAR(2,763) and the Lasso-SS(2,3123) models at lags 1 and 2, respectively. The color of each circle shows the value of the corresponding AR coefficient estimate. The solid black lines indicate grouping of the 46 states into 10 regions as used in the CDC influenza surveillance report.}
\label{googleFluTrends_2006to2010_matrixPlot_2stage_VAR_lassoSS_A1_A2}
\end{center}
\end{figure}

We compare the temporal dependence structures discovered by the three models, i.e., the VAR(2), the sVAR(2, 763) and the Lasso-SS(2,3123). Figure \ref{googleFluTrends_2006to2010_matrixPlot_2stage_VAR_lassoSS_A1_A2} displays the estimated AR coefficients from the three models at lags 1 and 2, respectively. To illustrate the possible spatial interpretation of the dependence structure, we group the 46 states into 10 regions as suggested in the CDC influenza surveillance report \footnote{The CDC 10-region division can be found at http://www.cdc.gov/flu/weekly/}, which is indicated by the solid black lines in Figure \ref{googleFluTrends_2006to2010_matrixPlot_2stage_VAR_lassoSS_A1_A2}. From panels (a), (c) and (e), we can see that the AR coefficient estimates on the diagonal of $\hat{A}_{1}$ are large and positive in all three models. This observation is reasonable since influenza activity from the previous week should be predictive of influenza activity of the current week within the same region. But panel (a) shows that this diagonal signal is diluted by the noisy off-diagonal AR estimates in the VAR(2) model. And except for this diagonal signal of $\hat{A}_{1}$, the other AR coefficient estimates in the VAR(2) model are noisy and hard to interpret at both lags 1 and 2. In contrast, the diagonal signal of $\hat{A}_{1}$ is most dominant in panel (c) of the 2-stage sVAR(2,763) model, in which lots of the off-diagonal AR coefficients are zero. Additionally, the overall interpretability of the sVAR(2,763) and the Lasso-SS(2,3123) models is much better than the VAR(2) model, since both models provide much cleaner descriptions of the temporal dependence structures and reveal some interesting patterns. For example, both the sVAR(2,763) and the Lasso-SS(2,3123) models discover the interdependence among the influenza activity of the 6 states in Region 1, i.e., (CT, MA, ME, NH, RI, VT), as indicated by the first block of states in panels (c), (d), (d) and (f). This within-region dependence is moderately positive at lag 1 and slightly negative at lag 2. In the sVAR(2,763) and the Lasso-SS(2,3123) models, we also observe the cross-region influence from Region 8 of (CO, MT, US) into Region 6 of (AR, LA, NM, OK, TX). In spite of their general resemblance, the Lasso-SS(2,3123) model contains many more non-zero AR coefficients than the sVAR(2,763) model. In fact, the Lasso-SS(2,3123) model has a large number of small (in absolute value) but non-zero AR coefficients, especially those at lag 2 as shown in panel (f).

The reduced complexity of sVAR models not only leads to better interpretability, but also improves forecast performance. To this point, we compare the out-of-sample forecast performance between the three models. We use the Google Flu Trends data between the week of July 10, 2011 and the week of December 25, 2011 ($T_{\mathrm{test}}=24$) as the test data. For the comparison, we compute two quantities: the first is the h-step-ahead forecast root mean squared error ($\mathrm{RMSE}$), which is defined as,
\begin{equation*}
\mathrm{RMSE}(h) = [K^{-1}(T_{\mathrm{test}}-h+1)^{-1}\displaystyle\sum_{k=1}^{K}\sum_{t=T}^{T+T_{\mathrm{test}}-h}(\hat{Y}_{t+h,k}-Y_{t+h,k})^2]^{\frac{1}{2}},
\end{equation*}
where $\hat{Y}_{t+h,k}$ is the h-step-ahead forecast of $Y_{t+h,k}$ for $k=1,\ldots,K$; the second is
the logarithmic score (LS), e.g., see \citet{Gneiting2007}, which is defined as,
\begin{equation*}
\mathrm{LS} = (T_{\mathrm{test}}-1)^{-1}\displaystyle\sum_{t=T+1}^{T+T_{\mathrm{test}}-1}-\log p_{t}(Y_{t}),
\end{equation*}
where $p_{t}(\cdot)$ is the probability density function of the forecast distribution. Table \ref{googleFluTrends_RMSE_score_sVAR2stage_VAR_LassoSS} summarizes the forecast $\mathrm{RMSE}$ for a forecast horizon $h=1, 2, 3$ and $4$ as well as the LS of each model. The sVAR(2,763) model fitted by the 2-stage approach has the smallest forecast $\mathrm{RMSE}$ among the three models, while the most saturated model, the VAR(2) model, has the worst out-of-sample forecast performance. The 2-stage approach gives the best forecast performance since it excludes many seemingly spurious AR coefficients from the sVAR(2,763) model. But the VAR(2) model contains a large number of spurious AR coefficients and their presence makes the out-of-sample forecast much less reliable. In addition, as seen from the last column of Table \ref{googleFluTrends_RMSE_score_sVAR2stage_VAR_LassoSS}, the LS rule also favors the sVAR(2,763) model among the three. \\
\begin{table}
\begin{center}
\begin{tabular}{l|r|r|r|r|r}
Model                      & $h=1$ & $h=2$ & $h=3$ & $h=4$ & LS~~~\\
\hline
sVAR(2,763)              &315.5  & 337.8 & 374.4  & 420.9 & 305.2\\
\hline
Lasso-SS(2,3123)       &324.7 & 351.5 & 400.9 & 437.2 & 317.4 \\
\hline
VAR(2)                     &336.4  & 393.2  & 468.7  & 562.3 & 462.7\\
\hline
\end{tabular}
\caption{The h-step-ahead forecast root mean squared error ($\mathrm{RMSE}$) and the logarithmic score (LS) of the sVAR(2,763), the Lasso-SS(2,3123) and the VAR(2) models. The test period is from the week of July 10, 2011 to the week of December 25, 2011 ($T_{\mathrm{test}}=24$). The forecast horizon is $h=1, 2, 3$ and $4$.}
\label{googleFluTrends_RMSE_score_sVAR2stage_VAR_LassoSS}
\end{center}
\end{table}

{\bf Concentration levels of air pollutants}. In this application, we analyze a time series of concentration levels of four air pollutants, CO, NO, NO$_{2}$ , O$_{3}$, as well as the solar radiation intensity R. The data are recorded hourly during the year 2006 at Azusa, California and can be obtained from the Air Quality and Meteorological Information System (AQMIS). The time series for analysis is of dimension $K=5$ and with $T=8370$ observations. 
The same dataset was previously studied in \citet{Songsiri2009}. A similar dataset of the same 5 component series, but recorded at a different location, was analyzed in \citet{Dahlhaus2000, Eichler2006}. The methods employed in \citet{Dahlhaus2000, Eichler2006, Songsiri2009} are based on the {\em partial correlation graph model}, in which VAR models are estimated under sparsity constraints on the inverse spectrum of VAR processes. So the modeling interest of the partial correlation graph approach is sparsity in the frequency domain, i.e., zero constraints on the inverse spectrum, while our 2-stage approach is concerned about sparsity in the time domain, i.e., zero constraints on AR coefficients. For this example, we are interested in comparing the findings from the 2-stage sVAR model and the partial correlation graph model.

We apply the 2-stage approach to fit a sVAR model to the air pollution data. The pre-specified range of the autoregression order $p$ is $\mathbb{P}=\{0,1,2,\ldots,8\}$. The same range for $p$ was also used in \citet{Songsiri2009}. 
The first stage does not exclude any pair of marginal series and leads to a stage 1 model with $\tilde{p}=4$ and $\tilde{M}=10$, which contains $(5+2\times10)\times4=100$ non-zero AR coefficients. The second stage further refines the model and leads to a sVAR(4,64) model. The selection of the autoregression order $p^{*}=4$ coincides with the result in \citet{Songsiri2009}, which also used BIC for VAR order selection. However, the BIC value of the 2-stage sVAR(4,64) model is 15301 and it is lower than the best BIC value (15414) reported in Table 1.1 of \citet{Songsiri2009}. This is because the partial correlation graph approach used in \citet{Songsiri2009} is concerned about sparsity in the inverse spectrum rather than in the AR coefficients. So the AR coefficients estimated by the partial correlation graph approach are never exactly zero, and the resulted VAR model will contain spurious non-zeros. The presence of these spurious AR coefficients is one limitation of the partial correlation graph approach: such spurious non-zeros do not substantially increase the likelihood but inflate the BIC, and they also weaken the interpretability of fitted VAR models. Another limitation of the partial correlation graph approach is that it only deals with a small dimension, since in the partial correlation graph approach model selection is usually executed based on an exhaustive search of all possible patterns of sparsity constraints on the inverse spectrum, e.g., see \citet{Dahlhaus2000, Eichler2006, Songsiri2009}. The number of such patterns is $2^{K(K-1)/2}$, which reaches $2\times 10^{6}$ when $K=7$. Therefore the partial correlation graph approach is feasible only for a small dimension. In fact, the largest dimension of all numerical examples considered in \citet{Dahlhaus2000, Eichler2006, Songsiri2009} is 6. This is unlike our 2-stage approach, which is able to deal with higher dimensions, such as the $46$-dimensional process in the Google Flu Trends example.

\begin{figure}[!h]
\begin{center}
\includegraphics[width=0.98\textwidth]{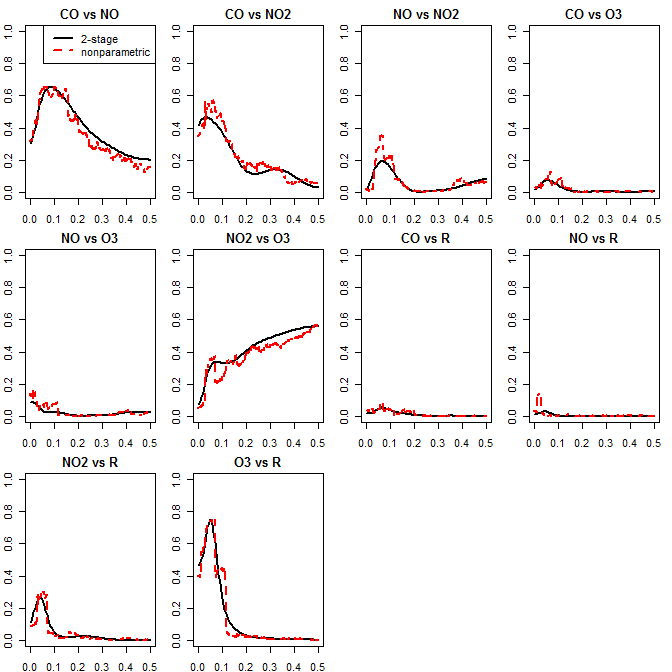}
\caption{Plots of the parametric estimates of the squared modulus of PSC, i.e., $|\mathrm{PSC}(\omega)|^2$, as computed from the AR coefficient estimates in the 2-stage sVAR(4,64) model (solid lines) and the non-parametric estimates of $|\mathrm{PSC}(\omega)|^2$ used in the first stage selection (dashed lines).}
\label{airPollution_ParCoh_2stage}
\end{center}
\end{figure}

Since the 2-stage approach is applied to the same dataset as in \citet{Songsiri2009}, it is interesting to compare the findings between the 2-stage sVAR model and the partial correlation graph model. Our comparison is in the frequency domain. Figure \ref{airPollution_ParCoh_2stage} displays the estimates of the squared modulus of PSC, i.e., $|\mathrm{PSC}(\omega)|^2$, as computed from the AR coefficient estimates in the 2-stage sVAR(4,64) model as well as the non-parametric estimates of $|\mathrm{PSC}(\omega)|^2$ used in the first stage of the 2-stage approach. We can see the good match-up between the two sets of estimates. So it is implied that it is possible to use the AR coefficient estimates from the 2-stage sVAR model, which are time-domain parameters, to recover the sparsity pattern in the inverse spectrum, which are frequency-domain quantities. We also point out that the estimates of $|\mathrm{PSC}(\omega)|^2$ from the 2-stage sVAR(4,64) model, as displayed in Figure \ref{airPollution_ParCoh_2stage}, resemble those in Figure 1.9 of \citet{Songsiri2009}, which displays the estimates of $|\mathrm{PSC}(\omega)|^2$ from the fitted partial correlation graph model. Furthermore, the findings from Figure \ref{airPollution_ParCoh_2stage} agree with the photochemical theory of interactions between the 5 marginal series. For example, the large estimates of $|\mathrm{PSC}(\omega)|^2$ between (CO, NO) comes from the fact that both air pollutants are mainly emitted from cars; the large estimates of $|\mathrm{PSC}(\omega)|^2$ between (O$_{3}$, R) reflects the major role of the solar radiation intensity in the generation of ozone, e.g., see \citet{Dahlhaus2000}. Additionally, from Figure \ref{airPollution_ParCoh_2stage} we observe that the estimates of $|\mathrm{PSC}(\omega)|^2$ between the pairs (CO, O$_{3}$), (CO, R),  (NO, R) and (NO, O$_{3}$) are relatively small as compared to the other pairs. This discovery of weak estimates of $|\mathrm{PSC}(\omega)|^2$ agrees with the findings in \citet{Dahlhaus2000, Eichler2006, Songsiri2009}, which are summarized in Table \ref{airPollution_weakPSC_summary}. For more detailed discussion on the underlying photochemical mechanism of interactions between air pollutants, readers are referred to \citet{Dahlhaus2000}.

\begin{table}
\begin{center}
\begin{tabular}{l|l}
 Model                     & Pairs with small estimates of $|\mathrm{PSC}(\omega)|^2$\\
\hline
2-stage sVAR(4,64)    &  (CO, O$_{3}$),  (CO, R), (NO, R), (NO, O$_{3}$)\\
\hline
\citet{Dahlhaus2000}  & (CO, O$_{3}$), (CO, R), (NO, R), (NO, O$_{3}$), (NO, NO$_{2}$)\\
\hline
\citet{Eichler2006} &  (CO, O$_{3}$),  (CO, R), (NO, R), (NO, O$_{3}$)\\
\hline
Songsiri et al.\cite{Songsiri2009} & (CO, O$_{3}$), (CO, R), (NO, R) \\
\hline
\end{tabular}
\caption{Pairs with weak estimates of $|\mathrm{PSC}(\omega)|^2$ in the 2-stage sVAR(4,64) model, as well as those found in \citet{Dahlhaus2000}, \citet{Eichler2006} and \citet{Songsiri2009}. \citet{Songsiri2009} used the same dataset as the sVAR(4,64) model; \citet{Dahlhaus2000} and \citet{Eichler2006} studied a similar dataset with the same 5 component series.}
\label{airPollution_weakPSC_summary}
\end{center}
\end{table}

\section{Discussion and Conclusion}\label{section_discussion_conclusion}
In this paper, we propose a 2-stage approach of fitting sVAR models, in which may of the AR coefficients are zero. The first stage of the approach is based on PSC and BIC to select non-zero AR coefficients. The combination of PSC and BIC provides an effective initial selection tool to determine the sparsity constraint on the AR coefficients. The second stage follows using $t$-ratios together with BIC to further refine the stage 1 model. The proposed approach is promising in that the 2-stage fitted sVAR models enjoy improved efficiency of parameter estimates and easier-to-interpret descriptions of temporal dependence, as compared to unrestricted VAR models. Simulation results show that the 2-stage approach outperforms Lasso-VAR methods in recovering the sparse temporal dependence structure of sVAR models. Applications of the 2-stage approach to two real data examples yield interesting findings about their temporal dynamics.

In the first stage selection of the 2-stage approach, we use \eqref{groupARzero} to link zero PSCs with zero AR coefficients. For some examples, however, this connection may not be exact. When non-zero AR coefficients correspond to zero PSCs, these AR coefficients are likely to be set to zero in the first stage and thus will not be selected by the 2-stage fitted models. For the cases we have investigated, however, we notice that purely BIC-selected models also tend to discard such AR coefficients. A possible explanation is that if the PSCs are near zero, the corresponding AR coefficients do not increase the likelihood sufficiently to merit their inclusion into the model based on BIC. As a result, the 2-stage approach still leads to sVAR models that perform similarly as the best BIC-selected models. To illustrate this point, we construct a VAR model in which a zero PSC corresponds to non-zero AR coefficients. Consider the following $3$-dimensional VAR(1) process $\{Y_{t}\}=\{(Y_{t,1},Y_{t,2},Y_{t,3})^{'}\}$ satisfying the recursions,
\begin{equation}\label{sim2equation}
\left(
\begin{array}{c}
Y_{t,1}\\
Y_{t,2}\\
Y_{t,3}\\
\end{array}
\right)
 =
\left(
\begin{array}{cccccc}
0 & 0.5  & 0.5  \\
0  & 0   & 0.3  \\
0  & 0.25  & 0.5 \\
\end{array}
\right)
\left(
\begin{array}{c}
Y_{t-1,1}\\
Y_{t-1,2}\\
Y_{t-1,3}\\
\end{array}
\right)
+
\left(
\begin{array}{c}
Z_{t,1}\\
Z_{t,2}\\
Z_{t,3}\\
\end{array}
\right),
\end{equation}
where $\{Z_{t}=(Z_{t,1},Z_{t,2},Z_{t,3})^{'}\}$ is iid Gaussian noise with mean {\bf 0} and covariance matrix,
\begin{equation*}
\Sigma_{Z} =
\left(
\begin{array}{cccccc}
 18  & 0 & 6 \\
  0  &  1 & 0 \\
  6  &  0 & 3 \\
\end{array}
\right).
\end{equation*}
For this example, one can show that $\mathrm{PSC}_{1,2}(\omega)=0$ for $\omega \in (-\pi, \pi]$ while $A_{1}(1,2)=0.5$. In applying the 2-stage approach to fit sVAR models to \eqref{sim2equation}, the first stage estimate of the summary statistic $\underset{\omega}{\operatorname{sup}} |\mathrm{PSC}_{1,2}(\omega)|^2$, as defined in \eqref{supPSCstat}, is likely to be small, so the estimates of $A_{1}(1,2)$ and $A_{1}(2,1)$ are likely to be automatically set to zero in the first stage.

\begin{figure}[p]
\begin{center}$
\begin{array}{cc}
\subfloat[number of non-zero AR coeff. estimates]{\includegraphics[width=0.48\textwidth]{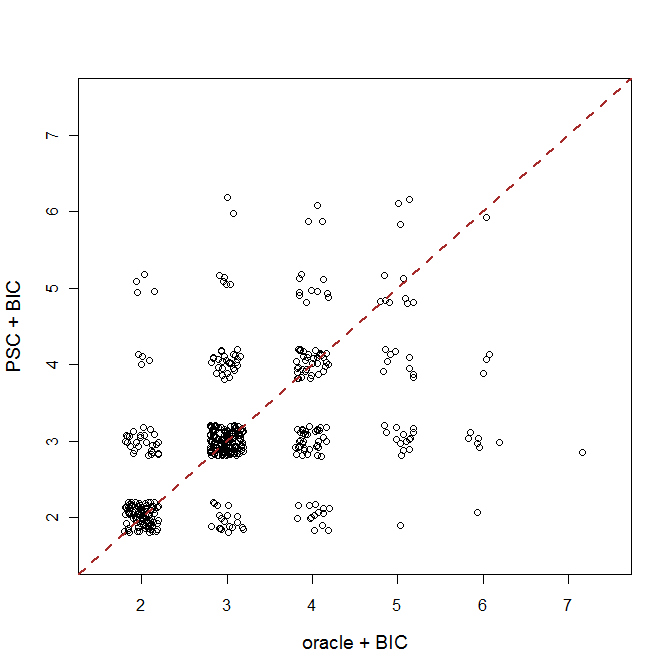}}  &
\subfloat[one-step forecast error]{\includegraphics[width=0.48\textwidth]{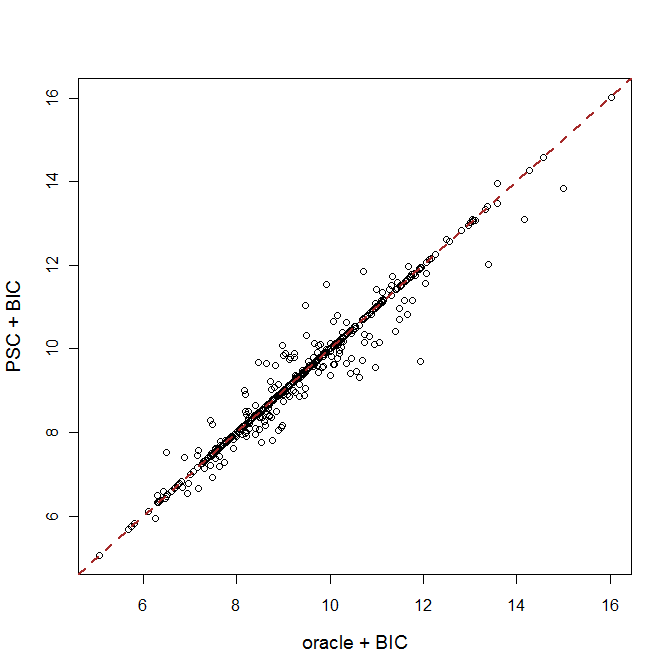}} \\
\subfloat[minus log-likelihood]{\includegraphics[width=0.48\textwidth]{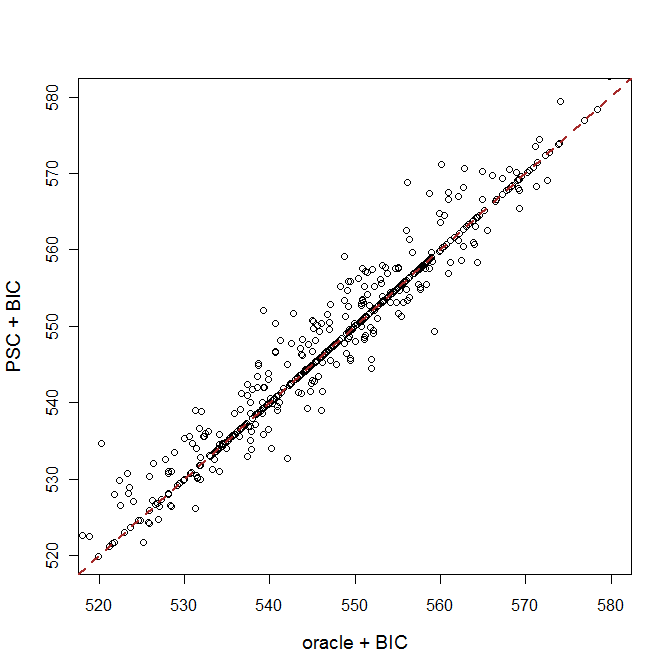}}  &
\subfloat[BIC]{\includegraphics[width=0.48\textwidth]{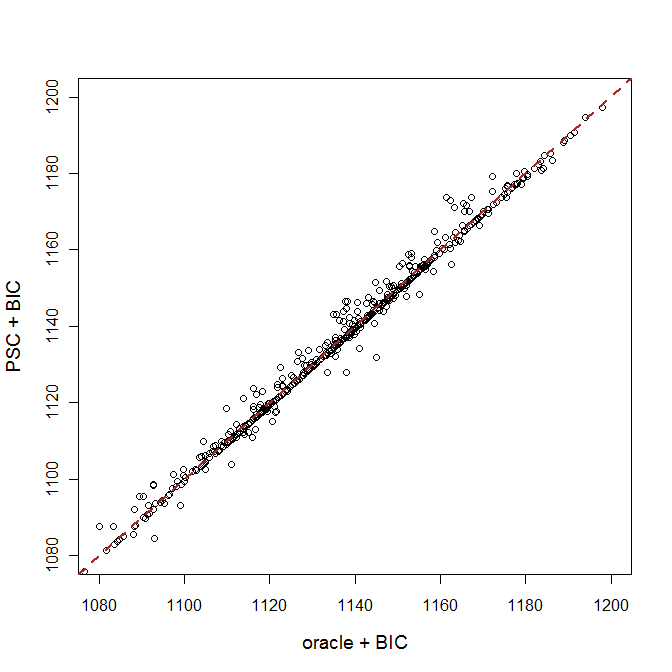}}
\end{array}$
\caption{Comparison between the 2-stage approach and the modified 2-stage procedure using different metrics. Panel (a): number of non-zero AR coefficient estimates. Panel (b): out-of-sample one-step forecast error. Panel (c): minus log-likelihood. Panel (d): BIC. In each panel, the x-axis refers to the modified 2-stage procedure and is labeled as ``oracle + BIC"; the y-axis refers to the original 2-stage approach and is labeled as ``PSC + BIC".}
\label{sim2_compare_2stage_specR}
\end{center}
\end{figure}

We compare the performance of the 2-stage approach with a modified 2-stage
procedure of fitting sVAR models to \eqref{sim2equation}. In the first stage of the modified procedure, we use precise knowledge of which AR coefficients are truly non-zero and conduct constrained maximum likelihood estimation under the corresponding parameter constraint. Then we execute the second stage of the modified procedure in exactly the same way as the original 2-stage approach. In other words, the modified procedure has an ``oracle" first stage and uses $t$-ratios together with BIC for further refinement in its second stage. So the truly non-zero AR coefficients will not be excluded after the first stage of the modified procedure. Such AR coefficients will survive the second stage refinement if the inclusion of them substantially increases the likelihood of the final sVAR model; otherwise they will be discarded after the second stage. For both approaches, the pre-specified range of the autoregression order $p$ is $\mathbb{P}=\{0,1,2,3\}$. The sample size $T$ is 100 and results are based on 500 replications. The comparison of these two approaches using different metrics is shown in Figure \ref{sim2_compare_2stage_specR}. In each panel of Figure \ref{sim2_compare_2stage_specR}, the x-axis refers to the modified 2-stage procedure and is labeled as``oracle + BIC"; the y-axis refers to the original 2-stage approach and is labeled as ``PSC + BIC". Panel (a) compares the number of non-zero AR coefficients, where these numbers are jittered so that their distributions can be observed; panel (b) compares the out-of-sample one-step forecast error; panel (c) compares the minus log-likelihood and panel (d) compares the BIC of the fitted models. From panel (a), we can see that the ``oracle + BIC" procedure does not lead to more non-zero AR coefficients than the 2-stage approach does. From panels (b), (c) and (d), we can see that the ``oracle + BIC" procedure does not provide improvement over the original 2-stage approach with respect to the one-step forecast error, the likelihood, or the BIC of fitted models. So, at least in this example, a non-zero AR coefficient that corresponds to a zero PSC is unlikely to be included in a BIC-selected model. As a result, our 2-stage approach has similar performance as that of the ``oracle + BIC" procedure. This phenomenon also raises the connection between the PSC and the likelihood of sVAR processes as an interesting direction for future research.

\newpage
\appendix
\section{Appendix}\label{section_appendix}
Appendix \ref{sVARest} gives results on the constrained maximum likelihood estimation of sVAR models. Appendix \ref{lassoToVAR} shows the procedure of implementing the two Lasso-VAR methods, i.e., the Lasso-SS and the Lasso-LL.

\subsection{Constrained maximum likelihood estimation of sVAR models}\label{sVARest}
Continuing with the notation in equation \eqref{VARequation}, the constraint that the AR coefficients of the VAR($p$) model are set to zero can be expressed as
\begin{equation}
\alpha\defeq\mathrm{vec}(A_{1},\ldots,A_{p}) = R\gamma, \label{sVARequation}
\end{equation}
where $\alpha = \mathrm{vec}(A_{1},\ldots,A_{p})$ is the $K^2p\times 1$ vector obtained by column stacking the AR coefficient matrices $A_{1},\ldots,A_{p}$; $R$ is a $K^2p\times m$ matrix of known constants with rank $m$ (usually $m \ll K^2p$); $\gamma$ is a $m\times 1$ vector of unknown parameters. The matrix $R$ in equation \eqref{sVARequation} is called the {\em constraint matrix} and it specifies which AR coefficients are set to zero by choosing one entry in each column to be $1$ and all the other entries in that column to be $0$. The rank $m$ of the constraint matrix $R$ equals the number of non-zero AR coefficients of the VAR model. This formulation is illustrated by the following simple example.

Consider a 2-dimensional zero-mean VAR(2) process $\{Y_{t}\}=\{(Y_{t,1},Y_{t,2})^{'}\}$ satisfying the recursions,
\begin{eqnarray}\label{sVARexample}
\left(
\begin{array}{c}
Y_{t,1} \\
Y_{t,2} \\
\end{array}
\right)
&= &
\left(
\begin{array}{cc}
A_{1}(1,1) & 0     \\
A_{1}(2,1) & A_{1}(2,2)\\
\end{array}
\right)
\times
\left(
\begin{array}{c}
Y_{t-1,1} \\
Y_{t-1,2} \\
\end{array}
\right) \\
&& +
\left(
\begin{array}{cc}
0 & 0     \\
A_{2}(2,1) & 0\\
\end{array}
\right)
\times
\left(
\begin{array}{c}
Y_{t-2,1} \\
Y_{t-2,2} \\
\end{array}
\right)
+
\left(
\begin{array}{c}
Z_{t,1} \\
Z_{t,2} \\
\end{array}
\right), \nonumber
\end{eqnarray}
where $A_{k}(i,j)$ is the $(i,j)$th entry of the AR coefficient matrix $A_{k}~(k=1,2)$. The VAR(2) model \eqref{sVARexample} contains 4 non-zero AR coefficients, $A_{1}(1,1), A_{1}(2,1), A_{1}(2,2)$ and $A_{2}(2,1)$, which can be expressed as
\begin{eqnarray}
\alpha &=& \mathrm{vec}(A_{1}, A_{2}) = R\gamma \nonumber \\
\Longrightarrow
\left(
\begin{array}{c}
A_{1}(1,1) \\
A_{1}(2,1) \\
0      \\
A_{1}(2,2) \\
0 \\
A_{2}(2,1) \\
0      \\
0 \\
\end{array}
\right)
&=&
\left(
\begin{array}{cccc}
1 & 0 & 0 & 0 \\
0 & 1 & 0 & 0 \\
0 & 0 & 0 & 0 \\
0 & 0 & 1 & 0 \\
0 & 0 & 0 & 0 \\
0 & 0 & 0 & 1 \\
0 & 0 & 0 & 0 \\
0 & 0 & 0 & 0 \\
\end{array}
\right)
\times
\left(
\begin{array}{c}
A_{1}(1,1)  \\
A_{1}(2,1)  \\
A_{1}(2,2)  \\
A_{2}(2,1)  \\
\end{array}
\right). \label{sVARequexa}
\end{eqnarray}
The constraint matrix $R$ in \eqref{sVARequexa} is of rank $m=4$, which equals to the number of non-zero AR coefficients. \\

\citet{Helmut1991} gives results on the constrained maximum likelihood estimation of the AR coefficients. Under the parameter constraint in the form of \eqref{sVARequation}, the maximum likelihood estimators of the AR coefficients $\alpha$ and the noise covariance matrix $\Sigma_{Z}$ are the solutions to the following equations,
\begin{eqnarray}
\hat{\alpha} &=& R\{R^{'}(LL^{'} \otimes \hat{\Sigma}_{Z}^{-1})R\}^{-1}R^{'}(L\otimes \hat{\Sigma}_{Z}^{-1})y, \label{sVARestAutoRegCoef} \\
\hat{\Sigma}_{Z} &=& \frac{1}{T-p}\displaystyle\sum_{t=p+1}^{T}(Y_{t}-\hat{Y}_{t})(Y_{t}-\hat{Y}_{t})^{'}, \label{sVARestCov}
\end{eqnarray}
where $\otimes$ is the \textit{Kronecker} product and
\begin{eqnarray*}
L_{t}      &\defeq& (Y_{t}, Y_{t-1},\ldots, Y_{t-p+1})^{'},\\
L &\defeq& (L_{0}, L_{1},\ldots, L_{T-1}),\\
y            &\defeq& \mathrm{vec}(Y) = \mathrm{vec}(Y_{1}, Y_{2},\ldots, Y_{T}), \\
\hat{Y}_{t}&\defeq& \displaystyle\sum_{k=1}^{p}\hat{A}_{k}Y_{t-k}.
\end{eqnarray*}
It is known that, e.g., see \citet{Helmut1991, Reinsel1997}, if there is no parameter constraint on the AR coefficients, i.e., $R=I_{K^2p}$ in \eqref{sVARequation}, then the maximum likelihood estimator of the AR coefficients does not involve the noise covariance matrix $\Sigma_{Z}$. From equation \eqref{sVARestAutoRegCoef}, however, we can see that the presence of the parameter constraint \eqref{sVARequation} makes the estimation of the AR coefficients commingled with the estimation of the covariance matrix $\Sigma_{Z}$. Therefore we iteratively update the estimators $\hat{\alpha}$ and $\hat{\Sigma}_{Z}$ according to equations \eqref{sVARestAutoRegCoef} and \eqref{sVARestCov}, until convergence, to obtain the constrained maximum likelihood estimator of the AR coefficients.

\subsection{Implementation of Lasso for VAR models}\label{lassoToVAR}
We give details of the two Lasso implementations of fitting VAR models, i.e., the Lasso-SS and Lasso-LL VAR models. Notice that the VAR($p$) model \eqref{VARequation} can be written in the following compact form,
\begin{equation}
y  = \mathrm{vec}(Y) = (L^{'}\otimes I_{K})\alpha + \mathrm{vec}(Z),  \label{VARcompact}
\end{equation}
where $\mathrm{vec}$ column stack operator, $\otimes$ is the \textit{Kronecker} product and
\begin{eqnarray*}
Y           &\defeq& (Y_{1}, Y_{2},\ldots, Y_{T}),\\
y            &\defeq& \mathrm{vec}(Y),\\
L_{t}      &\defeq& (Y_{t}, Y_{t-1},\ldots, Y_{t-p+1})^{'},\\
L            &\defeq& (L_{0}, L_{1},\ldots, L_{T-1}),\\
Z           &\defeq& (Z_{1}, Z_{2},\ldots, Z_{T}).
\end{eqnarray*}
Since $Z_{1},\ldots, Z_{T}$ are iid from the $K$-dimensional Gaussian $N(0, \Sigma_{Z})$, from \eqref{VARcompact} the minus log likelihood of the VAR($p$) model \eqref{VARcompact}, ignoring an additive constant, is,
\small
\begin{equation}
 -2\log L(\alpha, \Sigma_{Z}) = T\log|\Sigma_{Z}| + [y-(L^{'}\otimes I_{K})\alpha]^{'}(I_{T}\otimes \Sigma_{Z}^{-1})[y-(L^{T}\otimes I_{K})\alpha].
\end{equation}
\normalsize
For Lasso-penalized VAR models, there are two possible choices of the loss function: one is the sum of squared residuals and the other one is the minus log likelihood. The Lasso-SS method uses the sum of squared residuals as the loss function and the corresponding target function is,
\begin{equation}
Q^{SS}_{\lambda}(\alpha)\defeq ||y-(L^{'}\otimes I_{K})\alpha||_{2}^{2} + \lambda||\alpha||_{1}; \label{targetFunctionLassoSS}
\end{equation}
while the Lasso-LL method chooses the minus log likelihood as the loss function and its target function is,
\begin{eqnarray}\label{targetFunctionLassoLL}
~~~~~~~~ Q^{LL}_{\lambda}(\alpha, \Sigma_{Z}) \defeq
[y-(L^{'}\otimes I_{K})\alpha]^{'}(I_{T}\otimes \Sigma_{Z}^{-1})[y-(L^{'}\otimes I_{K})\alpha] \\
+ T\log|\Sigma_{Z}| + \lambda||\alpha||_{1}. ~~~~~~~~~~~~~~~~~~~~~~~~~~~~~~\nonumber
\end{eqnarray}
In both equations \eqref{targetFunctionLassoSS} and \eqref{targetFunctionLassoLL} the scalar tuning parameter $\lambda \in \mathbb{R}$ controls the amount of penalty. The AR coefficients $\alpha$ of the VAR model are estimated by minimizing the target function $Q^{SS}_{\lambda}(\alpha)$ \eqref{targetFunctionLassoSS} or $Q^{LL}_{\lambda}(\alpha, \Sigma_{Z})$ \eqref{targetFunctionLassoLL}, respectively.

It is worth noting that, unlike the linear regression model, the choice between the sum of squared residuals and minus log likelihood as the loss function will lead to different results of applying the Lasso method to VAR models. This can be seen by taking the first derivative of the Lasso-SS target function \eqref{targetFunctionLassoSS} and the Lasso-LL target function \eqref{targetFunctionLassoLL} with respect to the AR coefficient $\alpha$,
\begin{eqnarray}
\frac{\partial Q^{SS}_{\lambda}(\alpha)}{\partial \alpha} &=& 2[(LL^{'}\otimes I_{K}) - (L\otimes I_{K})y] + \lambda\cdot\mathrm{sgn}(\alpha), \label{LassoSSderivative}\\
\frac{\partial Q^{LL}_{\lambda}(\alpha)}{\partial \alpha} &=&  2[(LL^{'}\otimes\Sigma_{Z}^{-1}) - (L\otimes\Sigma_{Z}^{-1})y] + \lambda\cdot\mathrm{sgn}(\alpha), \label{LassoLLderivative}
\end{eqnarray}
where $\mathrm{sgn}(\cdot)$ is the {\em signum} function and $\mathrm{sgn}(\alpha)$ is the $K^2p\times 1$ vector in which the $k$th entry is $\mathrm{sgn}(\alpha_{k})$, $k=1,\ldots,K^2p$. We can see that noise covariance matrix $\Sigma_{Z}$ is taken into account by the Lasso-LL derivative \eqref{LassoLLderivative} but not by the Lasso-SS derivative \eqref{LassoSSderivative}. The two $K^2p\times 1$ vectors of first derivatives \eqref{LassoSSderivative} and \eqref{LassoLLderivative} are in general not equal (up to multiplication by a scalar) unless the covariance matrix $\Sigma_{Z}$ is a multiple of the identity matrix $I_{K}$. Therefore the Lasso-SS and the Lasso-LL methods will in general result in different VAR models.

Based on \eqref{targetFunctionLassoSS} and \eqref{targetFunctionLassoLL}, we describe the estimation procedures of the two Lasso-penalized VAR models. The estimation of Lasso-SS VAR models is straightforward since it can be viewed as standard linear regression problems with the Lasso penalty. Therefore the Lasso-SS VAR model can be fitted efficiently by applying the least angle regression (LARS) algorithm, e.g., see \citet{Efron2004} or the coordinate descent algorithm, e.g., see \citet{Friedman2010}. In this paper we use the coordinate descent algorithm implemented in the {\em R} package {\em glmnet} for fitting Lasso-SS VAR models. The estimation of Lasso-LL VAR models is more complicated since the target function \eqref{targetFunctionLassoLL} involves the unknown noise covariance matrix $\Sigma_{Z}$. We propose an iterative procedure to fit the Lasso-LL VAR model. The procedure is based on the fact that, for a given covariance matrix $\Sigma_{Z}$, the Lasso-LL target function \eqref{targetFunctionLassoLL} can be re-cast in a least-squares fashion. In other words, for a $K\times K$ positive-definite matrix $\Sigma_{Z}$, let
\begin{equation*}
\Sigma_{Z}=U\mathrm{diag}\{\kappa_{1},\ldots,\kappa_{K}\}U^{'},
\end{equation*}
be its eigenvalue decomposition, where $U$ is an orthonormal matrix and $\kappa_{1}\ge\kappa_{2}\ldots\ge\kappa_{K}>0$ are the $K$ positive eigenvalues. Define
\begin{equation}
\Sigma_{Z}^{-\frac{1}{2}}\defeq U\mathrm{diag}\{\frac{1}{\sqrt{\kappa_{1}}},\ldots,\frac{1}{\sqrt{\kappa_{K}}}\}U^{'} \label{invSqRootMatr}
\end{equation}
to be the {\em inverse square root} of $\Sigma_{Z}$. Notice that $\Sigma_{Z}^{-\frac{1}{2}}$ in \eqref{invSqRootMatr} is symmetric and $\Sigma_{Z}^{-\frac{1}{2}}\Sigma_{Z}^{-\frac{1}{2}}=\Sigma_{Z}^{-1}$, then we have
\begin{eqnarray*}
I_{T}\otimes \Sigma_{Z}^{-1} &=& (I_{T}\otimes \Sigma_{Z}^{-\frac{1}{2}}) (I_{T}\otimes \Sigma_{Z}^{-\frac{1}{2}}) \\
&=&  (I_{T}\otimes \Sigma_{Z}^{-\frac{1}{2}})^{'} (I_{T}\otimes \Sigma_{Z}^{-\frac{1}{2}}),\\
(I_{T}\otimes \Sigma_{Z}^{-\frac{1}{2}})[y-(L^{'}\otimes I_{K})\alpha] &=& (I_{T}\otimes \Sigma_{Z}^{-\frac{1}{2}})y - (I_{T}\otimes \Sigma_{Z}^{-\frac{1}{2}})(L^{'}\otimes I_{K})\alpha \\
&=& (I_{T}\otimes \Sigma_{Z}^{-\frac{1}{2}})y - (L^{'}\otimes \Sigma_{Z}^{-\frac{1}{2}})\alpha.
\end{eqnarray*}
Therefore the Lasso-LL target function \eqref{targetFunctionLassoLL} can be re-written as
\begin{align}\label{LassoLLtoLassoSS}
  &Q^{LL}_{\lambda}(\alpha, \Sigma_{Z})  \\
=&~T\log|\Sigma_{Z}| + [y-(L^{'}\otimes I_{K})\alpha]^{'}(I_{T}\otimes \Sigma_{Z}^{-1})[y-(L^{'}\otimes I_{K})\alpha] + \lambda||\alpha||_{1} \nonumber \\
=&~T\log|\Sigma_{Z}| + [y-(L^{'}\otimes I_{K})\alpha]^{'}(I_{T}\otimes \Sigma_{Z}^{-\frac{1}{2}})^{'} (I_{T}\otimes \Sigma_{Z}^{-\frac{1}{2}})[y-(L^{'}\otimes I_{K})\alpha] + \lambda||\alpha||_{1} \nonumber \\
=&~T\log|\Sigma_{Z}| + [(I_{T}\otimes \Sigma_{Z}^{-\frac{1}{2}})y - (L^{'}\otimes \Sigma_{Z}^{-\frac{1}{2}})\alpha]^{'}[(I_{T}\otimes \Sigma_{Z}^{-\frac{1}{2}})y - (L^{'}\otimes \Sigma_{Z}^{-\frac{1}{2}})\alpha] + \lambda||\alpha||_{1} \nonumber \\
=&~T\log|\Sigma_{Z}| + ||(I_{T}\otimes \Sigma_{Z}^{-\frac{1}{2}})y - (L^{'}\otimes \Sigma_{Z}^{-\frac{1}{2}})\alpha||_{2}^{2} + \lambda||\alpha||_{1}. \nonumber
\end{align}
The loss function
\begin{equation*}
||(I_{T}\otimes \Sigma_{Z}^{-\frac{1}{2}})y - (L^{'}\otimes \Sigma_{Z}^{-\frac{1}{2}})\alpha||_{2}^{2},
\end{equation*}
in \eqref{LassoLLtoLassoSS} can be viewed as the sum of squared residuals from a linear regression model with the response variable being $(I_{T}\otimes \Sigma_{Z}^{-\frac{1}{2}})y$ and the explanatory variables given by $L^{'}\otimes \Sigma_{Z}^{-\frac{1}{2}}$. Therefore, for a given $\Sigma_{Z}$, minimizing the Lasso-LL target function \eqref{LassoLLtoLassoSS} with respect to the AR coefficients $\alpha$ is equivalent to minimizing a Lasso-SS target function corresponding to the response variable $(I_{T}\otimes \Sigma_{Z}^{-\frac{1}{2}})y$ and the explanatory variables $L^{'}\otimes \Sigma_{Z}^{-\frac{1}{2}}$. So we can use the following iterative procedure to fit Lasso-LL VAR models.

\begin{center}
\Ovalbox{\setlength{\itemsep}{0pt}
\begin{minipage}{0.97\textwidth}
\begin{center}
{\bf An iterative procedure of fitting Lasso-LL VAR models}
\end{center}
\begin{itemize}\label{iteLassoLL}
\item[1.] Set an initial value $\Sigma_{Z}^{(0)}$ for the covariance matrix $\Sigma_{Z}$.
\item[2.] Update the AR coefficients $\alpha$ and the covariance matrix $\Sigma_{Z}$ at the $(k+1)$th iteration, until convergence, as follows,
\begin{itemize}
\item[2.1.] $\alpha^{(k+1)} = \underset{\alpha}{\operatorname{argmin}}~Q^{LL}_{\lambda}(\alpha, \Sigma_{Z}^{(k)})$ by applying the coordinate \\descent algorithm;
\item[2.2.] $\Sigma_{Z}^{(k+1)} = \frac{1}{T}(Y-A^{(k+1)}L)(Y-A^{(k+1)}L)^{'}$, \\where $\alpha^{(k+1)}=\mathrm{vec}(A^{(k+1)})$.
\end{itemize}
\end{itemize}
\end{minipage}}
\end{center}

Fitting Lasso-penalized VAR models, as all penalized regression methods, also involves choosing the tuning parameter $\lambda \in \mathbb{R}$. The choice of $\lambda$ is usually based on certain information criterion or cross-validations. In this paper we use cross-validations to determine the value of $\lambda$. Furthermore, the number of explanatory variables, i.e., the number of lagged values appearing on the right hand side of equation \eqref{VARcompact}, also depends on the unknown order of autoregression $p$. Therefore the values of both $p$ and $\lambda$ need to be determined in a data-driven manner. Suppose the autoregression order $p$ is restricted to take values in a pre-specified range $\mathbb{P}$, we use the following steps to fit Lasso-SS as well as Lasso-LL VAR models.

\begin{center}
\Ovalbox{\setlength{\itemsep}{0pt}
\begin{minipage}{0.97\textwidth}
\begin{center}
{\bf Steps of fitting Lasso-SS and Lasso-LL VAR models}
\end{center}
\begin{itemize}
\item[1.] For each $p \in \mathbb{P}$, apply the coordinate descent algorithm to minimize the Lasso-SS target function \eqref{targetFunctionLassoSS} and the aforementioned iterative procedure to minimize the Lasso-LL target function \eqref{targetFunctionLassoLL}, respectively.
For either the Lasso-SS or the Lasso-LL model, the optimal tuning parameter $\lambda^{opt}(p)$, depending on the given autoregression order $p$, is determined by the minimum average ten-fold cross-validation error, which is denoted by $\mathrm{CV}_{min}(p)$.
\item[2.] Choose $p^{*}$ that gives the minimum average cross-validation error over $\mathbb{P}$ 
as the autoregression order for either the Lasso-SS or the Lasso-LL VAR model.
\item[3.]  Obtain either the Lasso-SS or the Lasso-LL VAR model by setting the autoregression order $p$ equal to $p^{*}$ and the tuning parameter $\lambda$ equal to $\lambda^{opt}(p^{*})$.
\end{itemize}
\end{minipage}}
\end{center}

\section*{Acknowledgements}
We would like to thank Professor Songsiri for providing the air pollutant data. The research of Richard A. Davis is supported in part by the National Science Foundation grant DMS-1107031. The research of Tian Zheng is, in parts, supported by NSF grant SES-1023176 and a 2010 Google research award.

\bibliographystyle{asa}
\bibliography{paperForSparseVAR_arXiv_submission}

\begin{thebibliography}{31}
\newcommand{\enquote}[1]{``#1''}
\expandafter\ifx\csname natexlab\endcsname\relax\def\natexlab#1{#1}\fi

\bibitem[{Arnold et~al.(2008)Arnold, Liu, and Abe}]{Arnold2008}
Arnold, A., Liu, Y., and Abe, N. (2008), \enquote{Temporal causal modeling with
  graphical Granger methods,} \textit{Proceedings of the 13th ACM SIGKDD
  International Conference on Knowledge Discovery and Data Mining}.

\bibitem[{B{\"{o}}hm and von Sachs(2009)}]{Bohm2009}
B{\"{o}}hm, H. and von Sachs, R. (2009), \enquote{Shrinkage estimation in the
  frequency domain of multivariate time series,} \textit{Journal of
  Multivariate Analysis}, 100, 913--935.

\bibitem[{Brillinger(1981)}]{Brillinger1981}
Brillinger, D.~R. (1981), \textit{Time Series: Data Analysis and Theory}, New
  York: Holt, Rinehart and Winston.

\bibitem[{Brockwell and Davis(1991)}]{Davis1991}
Brockwell, P.~J. and Davis, R.~A. (1991), \textit{Time Series: Theory and
  Methods}, New York: Springer-Verlag.

\bibitem[{Dahlhaus(2000)}]{Dahlhaus2000}
Dahlhaus, R. (2000), \enquote{Graphical interaction models for multivariate
  time series,} \textit{Metrika}, 51, 157--172.

\bibitem[{Dahlhaus et~al.(1997)Dahlhaus, Eichler, and
  Sandk{\"{u}}hler}]{Dahlhaus1997}
Dahlhaus, R., Eichler, M., and Sandk{\"{u}}hler, J. (1997),
  \enquote{Identification of synaptic connections in neural ensembles by
  graphical models,} \textit{Journal of Neuroscience Methods}, 77, 93--107.

\bibitem[{Dempster(1972)}]{Dempster1972}
Dempster, A.~P. (1972), \enquote{Covariance selection,} \textit{Biometrics},
  28, 157--175.

\bibitem[{Duki{\'{c}} et~al.(2010)Duki{\'{c}}, Lopes, and Polson}]{Lopes2010}
Duki{\'{c}}, V., Lopes, H.~F., and Polson, N.~G. (2010), \enquote{Tracking flu
  epidemics using Google flu trends and particle learning,} \textit{Working
  paper}.

\bibitem[{Efron et~al.(2004)Efron, Hastie, Johnstone, and
  Tibshirani}]{Efron2004}
Efron, B., Hastie, T., Johnstone, T., and Tibshirani, R. (2004), \enquote{Least
  angle regression,} \textit{Annals of Statistics}, 32, 408--451.

\bibitem[{Eichler(2006)}]{Eichler2006}
Eichler, M. (2006), \enquote{Fitting graphical interaction models to
  multivariate time series,} \textit{Proceedings of the 22nd Conference on
  Uncertainty in Artificial Intelligence}.

\bibitem[{Eysenbach(2009)}]{Eysenbach2009}
Eysenbach, G. (2009), \enquote{Infodemiology: tracking flu-related searches on
  the web for syndromic surveillance,} \textit{AMIA: Annual Symposium
  Proceedings}, 244--248.

\bibitem[{Fan and Li(2001)}]{Fan2001}
Fan, J. and Li, R. (2001), \enquote{Variable selection via nonconcave penalized
  likelihood and its oracle properties,} \textit{Journal of the American
  Statistical Association}, 1348--1360.

\bibitem[{Fox and Dunson(2011)}]{Fox2011}
Fox, E. and Dunson, D. (2011), \enquote{Bayesian nonparametric covariance
  regression,} \textit{Arxiv preprint arXiv:1101.2017}.

\bibitem[{Friedman et~al.(2008)Friedman, Hastie, and Tibshirani}]{Friedman2008}
Friedman, J., Hastie, T., and Tibshirani, R. (2008), \enquote{Sparse inverse
  covariance estimation with the graphical lasso,} \textit{Biostatistics}, 9,
  432--441.

\bibitem[{Friedman et~al.(2010)Friedman, Hastie, and Tibshirani}]{Friedman2010}
--- (2010), \enquote{Regularization paths for generalized linear models via
  coordinate descent,} \textit{Journal of Statistical Software}, 33, 1--22.

\bibitem[{Ginsberg et~al.(2009)Ginsberg, Mohebbi, Patel, Brammer, Smolinski,
  and Brilliant}]{Ginsberg2009}
Ginsberg, J., Mohebbi, M., Patel, R., Brammer, L., Smolinski, M., and
  Brilliant, L. (2009), \enquote{Detecting influenza epidemics using search
  engine query data,} \textit{Nature}, 457, 1012--1014.

\bibitem[{Gneiting and Raftery(2007)}]{Gneiting2007}
Gneiting, T. and Raftery, A.~E. (2007), \enquote{Strictly proper scoring rules,
  prediction, and estimation,} \textit{Journal of the American Statistical
  Association}, 102, 359--378.

\bibitem[{Granger(1969)}]{Granger1969}
Granger, C. W.~J. (1969), \enquote{Investigating causal relations by
  econometric models and cross-spectral methods,} \textit{Econometrica}, 37,
  424--438.

\bibitem[{Haufe et~al.(2010)Haufe, M{\"{u}}ller, Nolte, and
  Kr{\"{a}}mer}]{Haufe2010}
Haufe, S., M{\"{u}}ller, K.~R., Nolte, G., and Kr{\"{a}}mer, N. (2010),
  \enquote{Sparse causal discovery in multivariate time series,}
  \textit{Journal of Machine Learning Research: Workshop and Conference
  Proceedings}, 6, 97--106.

\bibitem[{Hsu et~al.(2008)Hsu, Hung, and Chang}]{Hsu2008}
Hsu, N., Hung, H., and Chang, Y. (2008), \enquote{Subset selection for vector
  autoregressive processes using Lasso,} \textit{Computational Statistics and
  Data Analysis}, 52, 3645--3657.

\bibitem[{Hulth et~al.(2009)Hulth, Rydevik, and Linde}]{Hulth2009}
Hulth, A., Rydevik, G., and Linde, A. (2009), \enquote{Web queries as a source
  for syndromic surveillance,} \textit{PLoS ONE}, 4.

\bibitem[{Lozano et~al.(2009)Lozano, Abe, Liu, and Rosset}]{Lozano2009}
Lozano, A.~C., Abe, N., Liu, Y., and Rosset, S. (2009), \enquote{Grouped
  graphical Granger modeling for gene expression regulatory networks
  discovery,} \textit{Bioinformatics}, 25, 110--118.

\bibitem[{L{\"{u}}tkepohl(1993)}]{Helmut1991}
L{\"{u}}tkepohl, H. (1993), \textit{Introduction to Multiple Time Series
  Analysis}, New York: Springer-Verlag.

\bibitem[{Polgreen et~al.(2008)Polgreen, Chen, Pennock, and
  Forrest}]{Polgreen2008}
Polgreen, P.~M., Chen, Y., Pennock, D.~M., and Forrest, N.~D. (2008),
  \enquote{Using internet searches for influenza surveillance,}
  \textit{Clinical Infectious Diseases}, 47, 1443--1448.

\bibitem[{Reinsel(1997)}]{Reinsel1997}
Reinsel, G.~C. (1997), \textit{Elements of Multivariate Time Series Analysis},
  New York: Springer.

\bibitem[{Schwarz(1978)}]{Schwarz1978}
Schwarz, G. (1978), \enquote{Estimating the dimension of a model,}
  \textit{Annals of Statistics}, 6, 461--464.

\bibitem[{Shojaie and Michailidis(2010)}]{Michailidis2010}
Shojaie, A. and Michailidis, G. (2010), \enquote{Discovering graphical Granger
  causality using the truncating lasso penalty,} \textit{Bioinformatics}, 26,
  517--523.

\bibitem[{Song and Bickel(2011)}]{Song2011}
Song, S. and Bickel, P.~J. (2011), \enquote{Large vector auto regressions,}
  \textit{Arxiv preprint arXiv:1106.3915}.

\bibitem[{Songsiri et~al.(2010)Songsiri, Dahl, and Vandenberghe}]{Songsiri2009}
Songsiri, J., Dahl, J., and Vandenberghe, L. (2010), \enquote{Graphical models
  of autoregressive processes,} \textit{Convex Optimization in Signal
  Processing and Communications}, 89--116.

\bibitem[{Tibshirani(1996)}]{Tibshirani1996}
Tibshirani, R. (1996), \enquote{Regression shrinkage and selection via the
  Lasso,} \textit{Journal of the Royal Statistical Society, Series B}, 58,
  267--288.

\bibitem[{Vald{\'{e}}s-Sosa et~al.(2005)Vald{\'{e}}s-Sosa,
  S{\'{a}}nchez-Bornot, Lage-Castellanos, Vega-Hern{\'{a}}ndez, Bosch-Bayard,
  Melie-Garc{\'{i}}a, and Canales-Rodr{\'{i}}guez}]{Sosa2005}
Vald{\'{e}}s-Sosa, P.~A., S{\'{a}}nchez-Bornot, J.~M., Lage-Castellanos, A.,
  Vega-Hern{\'{a}}ndez, M., Bosch-Bayard, J., Melie-Garc{\'{i}}a, L., and
  Canales-Rodr{\'{i}}guez, E. (2005), \enquote{Estimating brain functional
  connectivity with sparse multivariate autoregression,} \textit{Philosophical
  Transactions of the Royal Society B}, 360, 969--981.

\end{thebibliography}

\end{document}